\documentclass[twocolumn, twocolappendix]{aastex63}

\usepackage{amsmath}
\usepackage{xspace}
\usepackage{outlines}
\usepackage{float}
\usepackage[bottom]{footmisc}
\usepackage{xcolor}
\usepackage[para,online,flushleft]{threeparttable}
\usepackage{tabularx}

\newcommand{\cotwo}{CO 2-1\xspace}
\newcommand{\velu}{km s$^{-1}$\xspace}

\newcommand{\voldensu}{$\mathrm{cm^{-3}}$\xspace}
\newcommand{\alphavir}{$\alpha_{\mathrm{vir}}$\xspace}
\newcommand{\alphaco}{$\alpha_{\mathrm{CO}}$\xspace}
\newcommand{\alphacou}{$\mathrm{M_{\odot}\ (K\ km\ s^{-1}\ pc^{2})^{-1}}$\xspace}
\newcommand{\solarmass}{M\ensuremath{_{\odot}}\xspace}
\newcommand{\Sigmol}{$\Sigma_{\mathrm{mol}}$\xspace}
\newcommand{\SigSFR}{$\Sigma_{\mathrm{SFR}}$\xspace}
\newcommand{\coldenunit}{M$_{\odot}$ pc$^{-2}$\xspace}
\newcommand{\vdep}{$\sigma_v$\xspace}
\newcommand{\fmol}{$f_{\mathrm{mol}}$\xspace}
\newcommand{\fgas}{$f_{\mathrm{gas}}$\xspace}
\newcommand{\Mmol}{$M_{\mathrm{mol}}$\xspace}
\newcommand{\Mgas}{$M_{\mathrm{gas}}$\xspace}
\newcommand{\Mstar}{$M_{\star}$\xspace}
\newcommand{\sfru}{$\mathrm{M_{\odot} yr^{-1}}$\xspace}
\newcommand{\tdep}{$t_{\mathrm{dep}}$\xspace}

\received{\today}
\revised{xxx}
\accepted{xxx}

\submitjournal{ApJ Letter}

\shorttitle{GMCs in Galaxy Mergers}
\shortauthors{He et al.}

\begin{document}

\title{Molecular gas and star formation in nearby starburst galaxy mergers}

\correspondingauthor{Hao He}
\email{heh15@mcmaster.ca}

\author[0000-0001-9020-1858]{Hao He}
\affiliation{McMaster University \\
1280 Main St W, Hamilton, ON L8S 4L8, CAN}

\author[0000-0003-4758-4501]{Connor Bottrell}
\affiliation{Kavli Institute for the Physics and Mathematics of the Universe (WPI), UTIAS, University of Tokyo\\ 
	 Kashiwa, Chiba 277-8583, Japan}

\author[0000-0001-5817-0991]{Christine Wilson}
\affiliation{McMaster University \\
	1280 Main St W, Hamilton, ON L8S 4L8, CAN}

\author[0000-0002-3430-3232]{Jorge Moreno}
\affiliation{Department of Physics and Astronomy, Pomona College, \\
			Claremont, CA 91711, USA}

\author[0000-0001-5817-5944]{Blakesley Burkhart} 
\affiliation{Department of Physics and Astronomy, 
Rutgers University, \\
136 Frelinghuysen Rd., 
Piscataway, NJ 08854, USA}
\affiliation{Center for Computational Astrophysics, 
Flatiron Institute, \\
162 Fifth Avenue, 
New York, NY 10010, USA}

\author[0000-0003-4073-3236]{Christopher C. Hayward}
\affiliation{Center for Computational Astrophysics, 
Flatiron Institute, \\
162 Fifth Avenue, 
New York, NY 10010, USA}

\author[0000-0001-6950-1629]{Lars Hernquist}
\affiliation{Center for Astrophysics, Harvard \& Smithsonian, \\
60 Garden Street, Cambridge, MA 02138, USA}   

\author{Angela Twum}
\affiliation{Department of Physics and Astronomy, Pomona College, \\
            Claremont, CA 91711, USA}

\begin{abstract}

We employ the Feedback In Realistic Environments (FIRE-2) physics model to study how the properties of giant molecular clouds (GMCs) evolve during galaxy mergers. We conduct a pixel-by-pixel analysis of molecular gas properties in both the simulated control galaxies and galaxy major mergers. The simulated GMC-pixels in the control galaxies follow a similar trend in a diagram of velocity dispersion ($\sigma_v$) versus gas surface density ($\Sigma_{\mathrm{mol}}$) to the one observed in local spiral galaxies in the Physics at High Angular resolution in Nearby GalaxieS (PHANGS) survey. For GMC-pixels in simulated mergers, we see a significant increase of factor of 5 -- 10 in both $\Sigma_{\mathrm{mol}}$ and $\sigma_v$, which puts these pixels above the trend of PHANGS galaxies in the $\sigma_v$ vs $\Sigma_{\mathrm{mol}}$ diagram. This deviation may indicate that GMCs in the simulated mergers are much less gravitationally bound compared with simulated control galaxies with virial parameter ($\alpha_{\mathrm{vir}}$) reaching  10 -- 100. Furthermore, we find that the increase in $\alpha_{\mathrm{vir}}$ happens at the same time as the increase in global star formation rate (SFR), which suggests stellar feedback is responsible for dispersing the gas. We also find that the gas depletion time is significantly lower for high \alphavir GMCs during a starburst event. This is in contrast to the simple physical picture that low \alphavir GMCs are easier to collapse and form stars on shorter depletion times. This might suggest that some other physical mechanisms besides self-gravity are helping the GMCs in starbursting mergers collapse and form stars.

\end{abstract}

.

\section{Introduction} \label{sec:intro}

Despite the diversity of galaxy morphology and environment, giant molecular clouds (GMCs) are 
the sites of star formation across cosmic time \citep{krumholz_star_2019, chevance2020}. As one of the most promising star formation model, the turbulence model \citep{krumholz_general_2005, hennebelle2011} suggest a relatively uniform star formation efficiency per freefall time ($\epsilon_{\mathrm{ff}}$) for individual GMCs. They predict that the observed scatter in $\epsilon_{\mathrm{ff}}$ could be account for by the diversity in GMC properties (e.g.virial parameter \alphavir and Mach number). However, \citet{lee_observational_2016} show that the observed scatter is larger than these early theoretical predictions expected and updated models suggest that cloud evolution, in addition to initial conditions such as Mach number and  \alphavir, should be accounted for \citep[see][]{burkhart2018,2018MNRAS.480.3916M,2019ApJ...879..129B}. Furthermore, \citet{grudic_when_2018} show in their simulation that GMCs in starburst galaxies can have different $\epsilon_{\mathrm{ff}}$ in normal spiral galaxies. Hence, to understand the links between GMCs and star formation in galaxies, it is essential to study various GMC properties in a broad range of environments.

However, modeling of GMCs starting from the scales of galaxies and cosmological zoom-ins is complicated by challenges in capturing the structure of the coldest and densest gas, which is heavily affected by various numerical choices, such as resolution \citep[e.g.][]{bournaud_high-resolution_2008,  teyssier_driving_2010} and the treatment of feedback \citep{fall_stellar_2010, murray_disruption_2010, dale_before_2014, myers_star_2014, raskutti_numerical_2016, kim_superbubbles_2017, grudic_when_2018,smith_efficient_2021}. Most resolved GMC simulations focus on the evolution of individual GMCs \citep[e.g.][]{burkhart2015a,howard_universal_2018, li_disruption_2019, decataldo_shaping_2020,burkhart2020} and ignore the wider environment. Only a handful of galaxy simulations have the ability to model GMC populations inside  Milky-Way-like galaxies \citep{jeffreson_general_2018, benincasa_live_2020} and mergers \citep{renaud_diversity_2019,li_formation_2022}.

High-resolution CO observations have successfully characterized GMCs in the Milky Way \citep[e.g.][]{rice_uniform_2016, rico-villas_super_2020, miville-deschenes_physical_2017, colombo_integrated_2019, lada_mass-size_2020} and nearby galaxies \citep[e.g.][]{donovan_meyer_resolved_2013,hughes_comparative_2013,colombo_pdbi_2014, leroy_portrait_2016, schruba_how_2019}. In particular, the recently completed PHANGS-ALMA survey \citep{leroy2021} has expanded these observations across a complete sample of nearby spiral galaxies, providing direct measurements of molecular gas surface density \Sigmol, velocity dispersion \vdep and size of GMCs, which are key quantities for determining the physical state of GMCs \citep[][]{larson_turbulence_1981}. Observations show that the correlation between $\sigma_v^2/R$ and \Sigmol is nearly linear \citep[e.g.,][]{heyer_molecular_2015, sun_cloud-scale_2018, sun_molecular_2020}, which is consistent with the theoretical prediction that most clouds follows the Larson's second law \citep{larson_turbulence_1981}, which indicates a constant ratio between clouds' kinetic energy and gravitational potential energy. This universal correlation provides us with a starting point to study how other galactic environmental factors (e.g., external pressure, stellar potential) influence the dynamical state of GMCs. 

%

Unlike studies targeting isolated galaxies, GMCs in starburst galaxy mergers are less well studied. On the observational side, the scarcity of nearby mergers means that we have only a handful of systems with GMC resolution data \citep{wei_two_2012, ueda2012,whitmore2014a,elmegreen2017, brunetti_highly_2020, brunetti_cloud-scale_2022, sanchez-garcia_duality_2022, bellocchi2022}. These studies show that GMCs in mergers have significantly higher gas surface densities and are less gravitationally bound compared to GMCs in normal spirals. However, it is difficult to draw statistically robust conclusions on how GMC properties evolve across various merging stages based on these limited number of local galaxy mergers. On the simulation front, only a handful of studies \citep[e.g.,][]{teyssier_driving_2010, renaud2014, fensch_high-redshift_2017} have the ability to probe the cold gas at $\sim$pc scale starting from cosmological scales. Using a comprehensive library of idealized galaxy merger simulations based on the FIRE-2 physics model,  \citet{moreno_interacting_2019} show that SFR enhancement is accompanied by an increase in the cold dense gas reservoir. This simulation suite thus provides us with the ideal tool to properly examine GMC evolution along the entire merging sequence. 

This paper explores how GMC properties evolve during the starburst merging event 
using the FIRE-2 merger suite from \citet{moreno_interacting_2019} and performs comparisons with observations to test the simulation model. In Section 2, we describe this simulation suite and the observational data used for comparison. Section 3 compares the  \vdep$-$\Sigmol relation between control simulated galaxies and normal spirals in the PHANGS-ALMA sample. Section 4 examines the \vdep$-$\Sigmol relation for mergers in both observations and simulations. In Section 5, we discuss and interpret various aspects of  the comparison between observations and simulations. 

\section{Data Processing}

\subsection{Simulated data}

\subsubsection{The FIRE-2 Model}

We use the FIRE-2 model \citep{hopkins_fire-2_2018}, which employs the hydrodynamic code GIZMO \citep{hopkins_new_2015, hopkins_new_2017}. Compared with the previous version, FIRE-2 adopts the updated meshless finite-mass (MFM) magnetohydrodynamics (MHD) solver, which is designed to capture the advantages of both grid-based and particle-based methods. We refer the reader to \citet{hopkins_new_2015} and \citet{hopkins_fire-2_2018} for details. The model includes treatment of radiative heating and cooling from free-free, photo-ionization/recombination, Compton, photoelectric, dust-collisional, cosmic ray, molecular, metal line, and fine-structure processes. Star formation occurs in gas that is self-gravitating (3D \alphavir $<1$ at the resolution scale), self-shielded, and denser than 1000 cm$^{-3}$ \citep[see Appendix C of][]{hopkins_fire-2_2018}.  Stellar feedback mechanisms include (i) mass, metal, energy, and momentum flux from supernovae type Ia \& II; (ii) continuous stellar mass-loss through OB/AGB winds; (iii) photoionization and photoelectric heating; and (iv) radiation pressure.  Each stellar particle is treated as a single stellar population. Mass, age, metallicity, luminosity, energy, mass-loss rate, and stellar feedback event rate for each stellar particle are calculated using the STARBURST99 stellar population synthesis model  \citep{leitherer_starburst99_1999}. The model does not account for feedback generated via accretion of gas onto a supermassive black hole (SMBH).

\subsubsection{Our FIRE-2 galaxy suite}
\label{subsec:merger_suite}

\citet{moreno_interacting_2019} present a suite of idealized galaxy merger simulations \citep[Initial conditions are manually set instead of from cosmological simulations; see also ][]{bottrell2019, moreno2021,mcelroy2022} covering a range of orbital parameters and mass ratios between 4 disc galaxies (G1, G2, G3 and G4, in order of increasing total stellar mass of (0.21, 1.24, 2.97 and 5.5$ \times 10^{10}$ \solarmass), along with separate runs for each disk galaxy in isolation (the control runs). Their orbit settings contain 3 orbital spin directions, 3 impact parameters and 3 impact velocities \citep[see Fig. 3 in ][]{moreno_interacting_2019}. For these simulations, the highest gas density and spatial resolution are 5.8 $\times$ 10$^5$ cm$^{-3}$ and 1.1 pc, respectively. The gravitational softening lengths are 10 pc for the dark matter and stellar components and 1 pc for the gaseous component. The mass resolution for a gas particle is 1.4 $\times 10^4$ \solarmass. The time resolution of a typical snapshot is 5 Myr \citep[See further details in][]{moreno_interacting_2019}.

\begin{table}[htb!]
    \centering
    \caption{Orbital Parameter of `e1' and `e2' orbit}
    \label{tab:orbital_info}
    \begin{threeparttable}
    \begin{tabularx}{0.3\textwidth}{lcc}
    \hline
    & e1 & e2 \\
    \hline
    Apo. Dist. (kpc)$^a$ & 60 & 120 \\
    Peri. Dist. (kpc)$^a$ & 15.5 & 9.3 \\
    \hline
    \end{tabularx}
    \begin{tablenotes}
    \item[a] First apocentric distance between the centers of two galaxies. \\
    \item[b] Second pericentric distance between the centers of two galaxies. 
    \end{tablenotes}
    \end{threeparttable}
\end{table}

For our analysis, we focus on the simulation run of isolated G2 and G3 galaxies along with one of G2\&G3 merger suites. The detailed information of G2 and G3 galaxies is in \citet[][Table 2]{moreno_interacting_2019}. The G2\&G3 merger suites have a mass ratio of 1:2.5 and hence are similar to major mergers such as the Antennae and NGC 3256 for which we have observational data. In addition, G2 and G3 have stellar masses within the range of the PHANGS sample \citep[10$^{10}$--10$^{11}$ \solarmass;][]{leroy2021}. We choose the `e' orbit \citep[][roughly prograde]{robertson2006}, which is expected to maximally enhance the star formation rate. In most of our analyses, we focus on the `e2' orbit since this is the fiducial run in \citet{moreno_interacting_2019}. We use the `e1' orbit as a comparison in some cases as it has smaller impact parameter and is more similar to the orbit of the Antennae merger \citep{privon_dynamical_2013}, for which we have GMC observational data. The pericentric distance of `e1' and `e2' orbit is listed in Table \ref{tab:orbital_info}. 

\subsubsection{Molecular gas}

We follow the scheme in \citet{moreno_interacting_2019} to separate the ISM of our simulated galaxy mergers into 4 components based on temperature and density: hot, warm, cool, and cold-dense gas, which roughly correspond to the hot, ionized, atomic, and molecular gas in observations \citep[see Table 4 in][]{moreno_interacting_2019}. The components that are most important for this work are the cool (temperatures below 8000 K and densities above 0.1 cm$^{-3}$) and the cold-dense gas (temperatures below 300 K and densities above 10 cm$^{-3}$), which corresponds to H I and H$_2$. This choice captures HI and H$_{\rm 2}$ gas reasonably well \citep{orr2018}. \citet{orr2018} also demonstrate that using this threshold to separate H$_2$ and HI yields reasonable agreement with the observed Kennicutt-Schmidt law \citep{kennicutt_global_1998, kennicutt_star_2012}. In the following, we refer to total gas as the sum of the gas in the cool and cold-dense phases (simulations) or in the atomic and molecular phases (observations).  

We adopt the same definition of molecular gas as in \citet{moreno_interacting_2019} (temperature below 300 K and density above 10 \voldensu). \citet{guszejnov_comparing_2017} demonstrate that the model successfully reproduces the GMC mass function in the Milky Way \citep{rice_uniform_2016} and the size-linewidth relation \citep[e.g., the Larson scaling relationship,][]{larson_turbulence_1981} in our Galaxy \citep{heyer_re-examining_2009, heyer_molecular_2015} and in nearby galaxies \citep[][]{bolatto2008, fukui_second_2008, muraoka_aste_2009,roman-duval_physical_2010,colombo_pdbi_2014,tosaki_statistical_2017}. Given the density cut of 10 \voldensu and mass resolution of 1.4 $\times 10^4$ \solarmass, the lower limit of our spatial resolution ($\sqrt[3]{M/\rho}$, where $M$ is the mass resolution and $\rho$ is the mass volume density) is $\sim$ 40 pc, which is smaller than the typical scale of observed GMCs \citep[40 -- 100 pc, ][]{rosolowsky_giant_2021}. In addition, GMC mass function peaks at 10$^5$ -- 10$^6$ \solarmass in Milky Way \citep{rice_uniform_2016}, which is significantly larger than our mass resolution. Therefore, we would generally expect more than 1 gas particle is included for molecular gas in each GMC-scale pixel. 

For generating different components of the ISM, the simulations start with a homogeneous ISM with a temperature of 10$^4$ K and solar metallicity. The multi-phase ISM then emerges quickly as a result of cooling and feedback from star formation. The initial gas mass for the simulation is set to match the median HI mass from the xCOLDGASS survey \citep{catinella_xgass_2018}. 

\subsubsection{Data cubes}

We first convert the FIRE-2 molecular gas data into mass-weighted position-position-velocity (p-p-v) data cubes to match the format of the CO data from radio observations \citep{mcmullin_casa_2007}. We adopt the cube construction method created for \citet{bottrell2022} and \citet{bottrell2022a} and then adapted to the FIRE-2 merger suite by \cite{mcelroy2022}.
Kinematic cubes are produced along four lines-of-sight (labeled as `v0', `v1', `v2', `v3'), defined by the vertices of a tetrahedron centered at the primary galaxy (G3 in this work).  For the isolated galaxy simulations, we generate p-p-v data cubes at different inclination angles (10 -- 80 degrees). We adopt a pixel size of 100 pc and velocity resolution of 2 km s$^{-1}$, which is similar to PHANGS choice \citep{sun_molecular_2020}. The field of view (FOV) for the data cube is set to be 25 kpc. 

Then we create zeroth-moment maps of the gas surface density \Sigmol and second-moment maps of the velocity dispersion \vdep. We do not set any thresholds on these moment maps since we argue that every gas particle in the simulated cube should be treated as a real signal, rather than observational noise. However, in later analyses, when we display \vdep versus \Sigmol for the simulated data, we select pixels with \Sigmol greater than 1 \coldenunit, which approximates the lower limit of the molecular gas detection threshold in the observational data \citep{sun_cloud-scale_2018}. We also exclude pixels detected in fewer than two velocity channels in the simulated cube to exclude inaccurate measurements of \vdep. 

To characterize clouds, we use a pixel-based analysis \citep{leroy_portrait_2016}, which treats each pixel as an individual GMC, rather than identifying each individual cloud from the data cube. This approach has been widely applied to GMC analyses for PHANGS galaxies \citep{sun_cloud-scale_2018, sun_molecular_2020}. Compared to the traditional cube-based approach, this new method requires minimal assumptions and can be easily applied to many datasets in a uniform way, while still giving us the essential GMC properties (e.g., molecular gas surface density \Sigmol, gas velocity dispersion \vdep). On the other hand, the pixel-based method has a major disadvantage of not able to decompose different cloud components along the same line of sight. Several observational studies \citep[e.g.][]{brunetti_extreme_2022, sun_molecular_2022} have compared this new approach with the traditional approach and found good agreement on cloud properties between two methods for both normal spiral galaxies and starburst mergers, especially for clouds in galaxy disks. These comparisons show pixel-based analysis should be valid for capturing individual cloud properties, especially for galaxy disks which generally have single-layer of GMCs (see Section \ref{subsec:alphaVir_SFR} for detailed discussion about the projection effect).  In this work, we adopt this approach to match the method in \citet{brunetti_highly_2020} and \citet{brunetti_cloud-scale_2022}. 
We also note that since we treat each pixel as a GMC, these GMCs do not necessarily represent independent ISM structures. In fact, given the mass resolution of 1.4 $\times 10^4$ \solarmass, we can barely resolve the internal structure of most massive GMCs of $\sim 10^6$ \solarmass (100 elements). We refer to them as GMCs in this paper to be consistent with similar observational analyses \citep[e.g.][]{sun_cloud-scale_2018, sun_molecular_2020}. 

\subsection{Star Formation Rate Maps}


To further explore how the GMC properties at 100 pc scale affect the star formation, we also make SFR maps with the same resolution of 100 pc for the simulated mergers at different times.  We create these maps using a method similar to the one used to create the gas cubes. We include all the stellar particles with age younger than 10 Myr and create p-p-v data cubes for these stellar particles. The mass-weighted cubes are integrated along the velocity axis to produce 2D maps of stellar mass formed within the last 10 Myr. These surface-density mass maps are subsequently divided by 10 Myr to obtain the average star-formation rates over the last 10 Myr.

\subsection{Observational Data}
We use several sets of observations for comparison with our simulations.
\subsubsection{Spiral galaxies: PHANGS data}

For isolated galaxies, we mainly use the PHANGS data from \citet{sun_molecular_2020} with resolution of 90 pc, which is comparable to our pixel size choice of 100 pc. \citet{sun_molecular_2020} apply the pixel-based method for statistical analyses of GMC properties for 70 galaxies in the PHANGS sample. We also include GMC data for M31 from \citet{sun_cloud-scale_2018} at resolution of 120 pc. M31 is identified as a green-valley galaxy, similar to our own Milky Way, and hence has a lower total gas fraction than normal spiral galaxies \citep{mutch_mid-life_2011}. Both M31 and the Milky Way seem to be in a transition from blue spiral galaxies to quenched galaxies via depletion of their cold gas (Bland-Hawthorn \& Gerhard 2016). M31 has stellar mass of 10$^{11}$ \solarmass \citep{sick_stellar_2015}, H$_2$ mass of $3.6 \times 10^8$ \solarmass and HI mass of $4.8 \times 10^9$ \solarmass \citep{nieten_molecular_2006}. 

\subsubsection{Galaxy mergers: the Antennae and NGC 3256}

\begin{table}[htb!]
    \centering
    \caption{Information about the observed mergers in this work}
    \begin{threeparttable}
    \movetableright=-0.5in
    \begin{tabular}{lccc}
    \hline
         & Antennae &  NGC 3256 & \#References \\
    \hline
    \Mstar ($10^{10}$ \solarmass)$^a$   & 4.5 & 11.4 & (1); (2) \\
    \Mmol ($10^{10}$ \solarmass)$^b$      & 1.2 & 0.8 & (3); this work \\
    SFR (\sfru)        & 8.5 & 50 & (1); (4) \\
    Sep. (kpc)$^c$   & 7.3 & 1.1 & (5); (4) \\
    $t_{\mathrm{now}}$ (Myr)$^d$  & 40 & $\cdots$ & (6) \\
    mass ratio$^f$ & 1:1 & $\cdots$ & (6) \\
    Peri. Sep (kpc)$^g$ & 10.4 & $\cdots$ & (6) \\
    \hline
    \end{tabular}
    \begin{tablenotes}
    \item Notes: \textit{a}. Stellar mass. \textit{b.} Molecular gas mass. \textit{c.} Current separation between two nuclei. \textit{d.} Current time since the second passage. \textit{e.} Mass ratio of the two progenitor galaxies. \textit{g.} Pericentric distance of two nuclei from the simulation model. \\
    \item References: (1) \citet{seille_spatial_2022} (2) \citet{howell2010}
    (3) \citet{wilson_highresolution_2000} (4) \citet{sakamoto_infrared-luminous_2014} (5) \citet{zhang_antennae_2001} (6) \citet{karl2010}
    \end{tablenotes}
    \end{threeparttable}
    \label{tab:obs_merger}
\end{table}

We use the \cotwo data for NGC 3256 \citep{brunetti_highly_2020} and the Antennae (\citet[][ Brunetti et al. in prep]{brunetti_cloud-scale_2022}) at resolutions of 90 and 80 pc, respectively. The GMC measurements use the same pixel-based approach as in \citet{sun_cloud-scale_2018, sun_molecular_2020}. Both NGC 3256 and the Antennae are identified as late-stage major mergers that have been through their second perigalactic passage \citep{privon_dynamical_2013}. NGC 3256 has stellar mass of 1.1 $\times$ 10$^{11}$ \solarmass, total molecular gas of 8 $\times$ 10$^{19}$ \solarmass (calculated based on \cotwo map in \citet{brunetti_extreme_2022}, assuming \alphaco of 1.1 \alphacou and \cotwo/1-0 ratio of 0.8) and SFR of 50 \sfru \citep{sakamoto_infrared-luminous_2014}. In contrast, the Antennae has a stellar mass of 4.5 $\times$ 10$^{10}$ \solarmass and SFR of 8.5 \sfru \citep{seille_spatial_2022}. NGC 3256 currently has a more intense starburst, perhaps because it is at different evolutionary stage in the merging process. The detailed information is in Table \ref{tab:obs_merger}. 

To convert the \cotwo emission to molecular gas mass requires the assumption of a CO-to-H$_2$ conversion factor (\alphaco). The exact value of \alphaco has large uncertainties and varies significantly among different types of galaxies, especially for starburst galaxies. \citet{downes_rotating_1998} find that for starburst U/LIRGs, the \alphaco value is generally 4 times smaller than that in our Milky Way. The major method for direct measurement of \alphaco is through large velocity gradient (LVG) radiative transfer modeling of multiple CO and its isotope lines. For \alphaco in the Antennae, various LVG modeling \citep[e.g.][]{zhu2003, schirm_herschel-spire_2014} suggests that the Antennae has \alphaco close to the Milky Way value of 4.3 \alphacou. This is also supported by the galaxy simulation that specifically matches the Antennae  \citep{renaud_diversity_2019}. 
For NGC 3256, we do not have a direct measurement of \alphaco. We therefore adopt the treatment from \citet[][]{sargent_regularity_2014} to determine the \alphaco for an individual galaxy as 
\begin{equation}
\alpha_{\mathrm{CO}} = (1-f_{\mathrm{SB}})\times \alpha_{\mathrm{CO, MS}} + f_{\mathrm{SB}} \times \alpha_{\mathrm{CO, SB}},  
\end{equation}
where $\alpha_{\mathrm{CO, MS}}$ and $\alpha_{\mathrm{CO, SB}}$ are the conversion factors for the Milky Way (4.3 \alphacou) and U/LIRGs (1.1 \alphacou, including helium), and $f_{\mathrm{SB}}$ is the probability for a galaxy to be a starburst galaxy, which is determined by its deviation from the star-forming main sequence. We adopt the star-forming main sequence relation from \citet{catinella_xgass_2018}, 
\begin{equation}
\log \mathrm{sSFR_{MS}} = -0.344(\log M_{\star} -9) - 9.822, 
\end{equation}
where sSFR = SFR / \Mstar~ is the specific star formation rate. NGC 3256 has an sSFR/$\mathrm{sSFR_{MS}}$ ratio of 15 \citep{brunetti_highly_2020}, which suggests NGC 3256 should have \alphaco close to the U/LIRG value of 1.1 \alphacou. Therefore, in the following analyses, we will adopt \alphaco of 4.3 \alphacou for the Antennae and 1.1 \alphacou for NGC 3256.



\section{Control (Isolated) Galaxies}
\label{sec:control}

\begin{figure}[htb!]
	\gridline{
		\fig{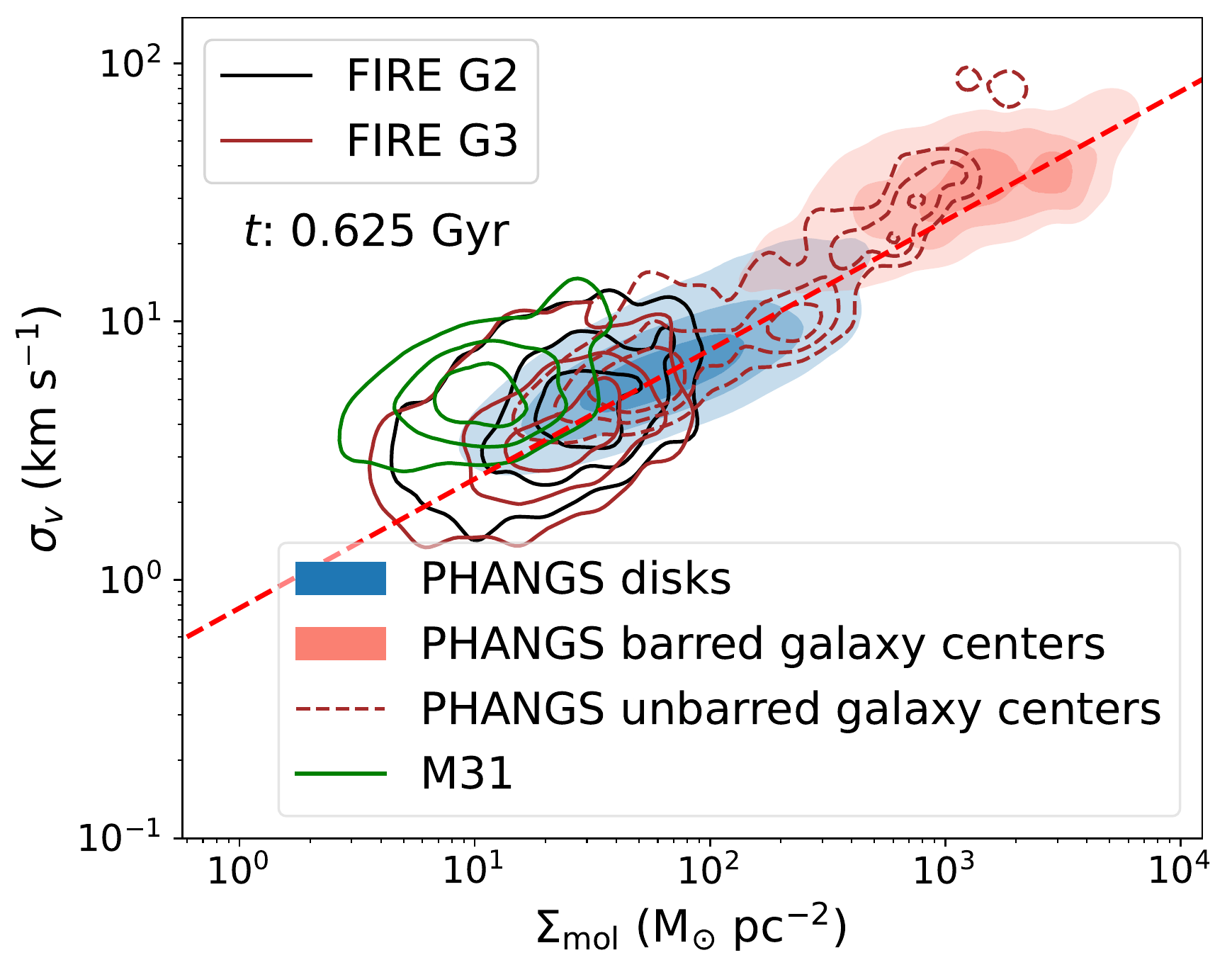}{0.45\textwidth}{}
	}
	\vspace{-0.2 in}
	\caption{Velocity dispersion versus gas surface density for the G2 (black solid contour) and G3 (brown solid contour) simulated galaxies at 0.625 Gyr with inclination angle of 30 degrees compared to the PHANGS galaxy sample. The contours are mass-weighted and set to include 20\%, 50\% and 80\% of the data. The density contours of PHANGS galaxies \citep{sun_molecular_2020} show the distribution of measurements in galaxy disks (blue shaded contours), the centers of barred galaxies (salmon shaded contours) and the centers of unbarred galaxies (brown dashed contours) with a resolution of 90 pc. The red dashed line marks the position of the median values of \alphavir for PHANGS galaxies of 3.1 \citep{sun_molecular_2020}. We also show the data for M31 (green solid contour) at 120 pc resolution from \citet{sun_cloud-scale_2018}. 
    \textit{We can see that the FIRE-2 spiral galaxies follow the same \vdep - \Sigmol relation as the PHANGS galaxies.} }
	\label{fig:Sigmol_vdep}
\end{figure}


To test if the simulation successfully reproduces observed GMCs, Figure~\ref{fig:Sigmol_vdep} shows the well-known correlation between \vdep and \Sigmol for isolated simulated galaxies and PHANGS-ALMA spiral galaxies. We show \vdep versus \Sigmol contours for G2 and G3 galaxies at an inclination angle of 30 degrees, compared with that of observed galaxies. The two simulated galaxies exhibit similar properties (black and dark red solid contours) and generally lie on the trend followed by the PHANGS galaxies. We also plot a red dashed line indicating GMCs with constant virial parameter \alphavir of 3.1. For the pixel-based analysis, \alphavir is calculated as \citep{sun_cloud-scale_2018}
\begin{equation}
\label{eq: alphavir}
\begin{split}
\alpha_{\mathrm{vir}} & = \frac{9 \ln2}{2 \pi G} \frac{\sigma_v^2 }{\Sigma_{\mathrm{mol}} R} \\
&= 5.77 \left(\frac{\sigma_v}{\mathrm{km\ s^{-1}}}\right)^2 \left(\frac{\Sigma_{\mathrm{mol}}}{\mathrm{M_{\odot}}}\right)^{-1} \left(\frac{R}{\mathrm{40 pc}}\right)^{-1}, 
\end{split}
\end{equation}
where $R$ is the GMC radius. In \citet{sun_cloud-scale_2018}, $R$ is set to be the radius of the beam in the image, as each beam is treated as an independent GMC. We can see both our simulated galaxies and observed PHANGS galaxies follow the trend of the constant \alphavir, which yields the relation of $\sigma_v^2 \propto \Sigma_{\mathrm{mol}}$ that suggests the simulations reproduce GMCs similar to the observations. However, we can see that the two galaxies lie at the low surface-density end of the PHANGS distribution and thus their gas properties are more similar to those of M31 than a typical PHANGS galaxy. Indeed, the molecular and total gas properties of the simulated galaxies are similar to those of M 31, perhaps due to the choice of initial gas mass in the simulations (see Appendix \ref{sec:gas_fraction}). 

\section{Merging galaxies}
\label{sec:merger}

\subsection{GMC linewidth and surface density}
\label{subsec:GMCs_mergers}

\begin{figure*}[htb!]
    \gridline{
        \fig{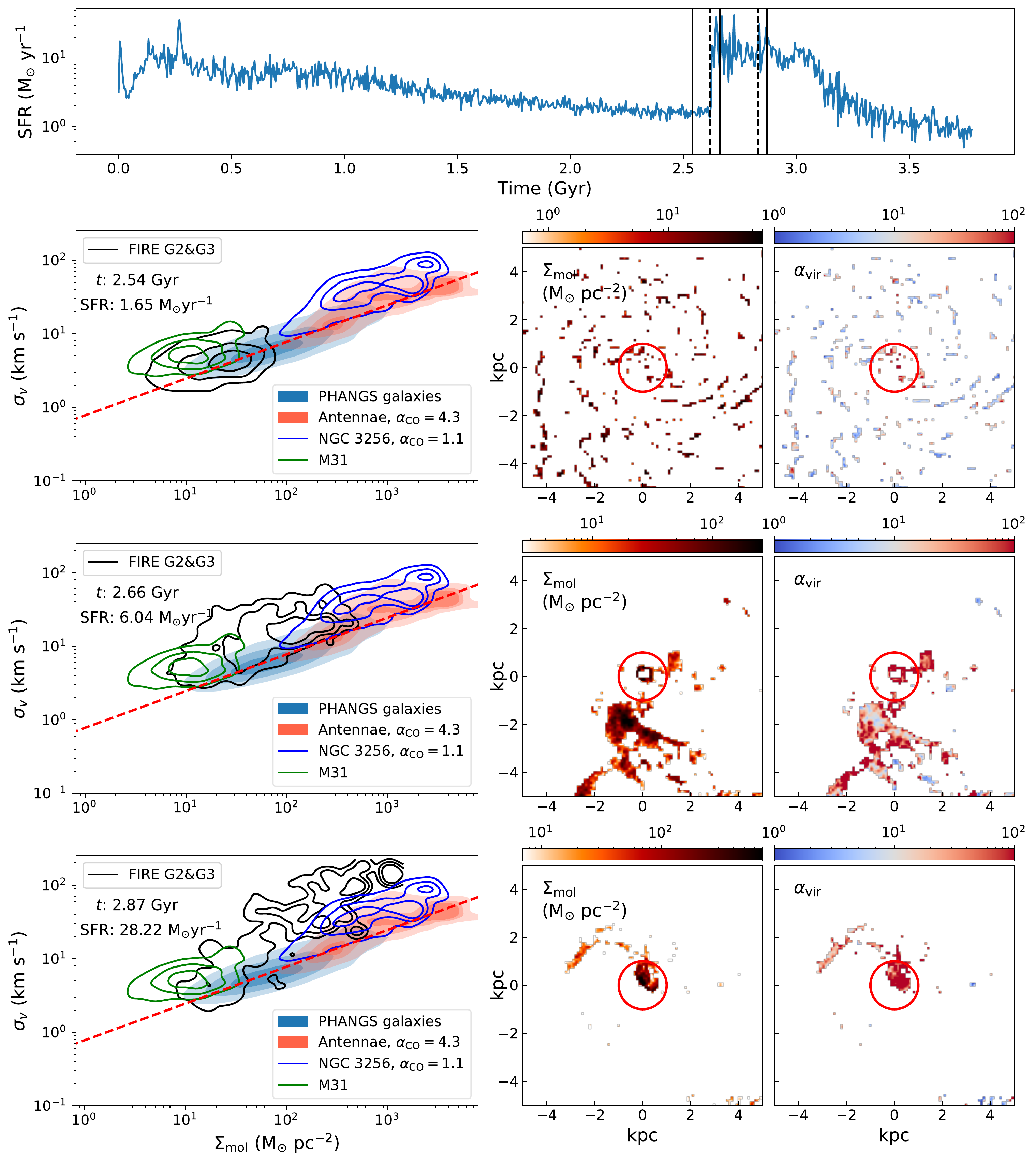}{0.95\textwidth}{}
        }
	\vspace{-0.2 in}
	\caption{(\textit{Top}) SFR history for the G2\&G3 merger with `e2' orbit with viewing angle of `v0'. The 3 solid black vertical lines indicate the time for each snapshot displayed below. The two dashed lines indicate the times at the start of second merging and the final coalesce of two nuclei. (\textit{Bottom}) Three snapshots. For each snapshot, the left panel shows  the \vdep versus \Sigmol mass-weighted contour with the same setting as Fig. \ref{fig:Sigmol_vdep}.  We also show the density contours for the PHANGS galaxies (filled blue region), NGC 3256 (blue contours) and the Antennae (orange shaded region). For NGC 3256, \Sigmol is calculated using the ULIRG \alphaco of 1.1 \alphacou. For the Antennae, the gas surface density is calculated using the Milky Way \alphaco of 4.3 \alphacou. The red dashed line indicate the line of constant \alphavir of 3.1. 
	The right two panels show the \Sigmol and \alphavir maps of inner 5 kpc regions where we have most of our detected pixels. The interactive version of the animation is available at \url{https://heh15.github.io/merger_animation/G2G3_e2_v0_final} that contains the panels of SFR history, \vdep vs \Sigmol diagram and \Sigmol map. You can drag the slide to view mergers at different times. We can see that the properties of the GMCs right before the second passage still resemble those of normal spiral galaxies, while GMCs after the second passage lie above the PHANGS trend in the \vdep vs \Sigmol plot and show significantly higher \alphavir. }
	\label{fig:Sigmol_vdep_merger1}
\end{figure*} 

\begin{figure*}[htb!]
    \gridline{
        \fig{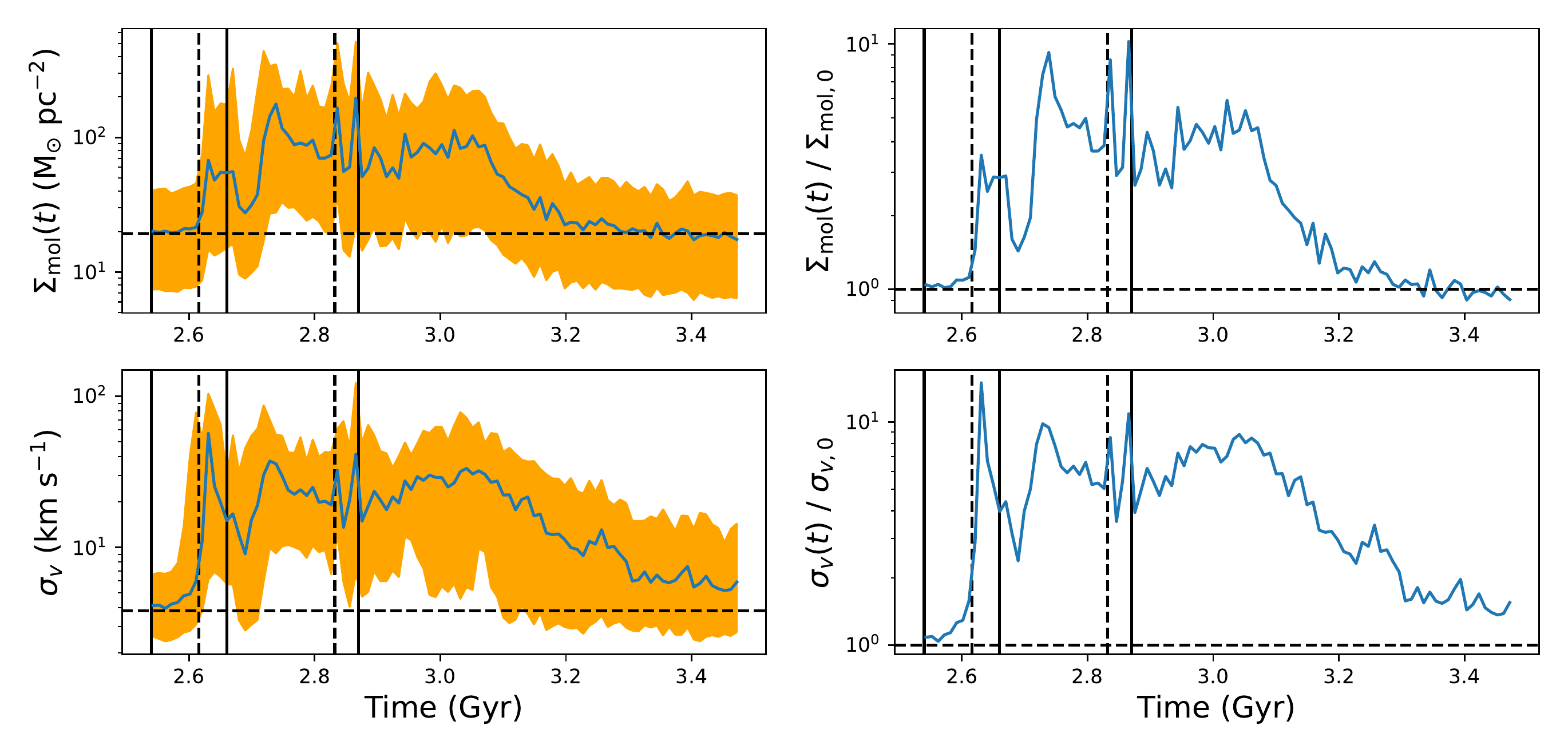}{0.9\textwidth}{}
        }
    \vspace{-0.4 in}
    \caption{The \Sigmol and \vdep variation across the second passage and final coalescence of the G2\&G3 merger at `e2' orbit with viewing angle of `v0'. The two dashed vertical lines indicate the times when the simulated merger begin the second passage and experience final coalescence. The three solid vertical lines correspond to the 3 snapshots shown in Fig, \ref{fig:Sigmol_vdep_merger1}. The horizontal dashed lines indicate the median value of the isolated G3 galaxy at time of 0.625 Gyr (Fig. \ref{fig:Sigmol_vdep}) as a baseline for comparison.  
    (\textit{Upper left}) \Sigmol vs time. Blue lines shows the mass weighted median \Sigmol of the entire merger while the orange filled area indicates \Sigmol range between 16th and 84th percentile. The two dashed lines indicate the time for the start of the second passage and the final coalesce of the two nuclei. (\textit{Upper right}) The ratio between median \Sigmol at given time and the median value $\Sigma_{\mathrm{mol},0}$ for the isolated G3 galaxy at 0.625 Gyr. (\textit{Lower left}) The mass-weighted median \vdep versus time. (\textit{Lower right}) The ratio between the median \vdep and the value $\sigma_{v,0}$ for isolated G3 galaxies at 0.625 Gyr. We can see both \Sigmol and \vdep increase dramatically during the second passage when the extreme starburst happens. }
    \label{fig:Sigmol_vdep_deviation}
\end{figure*}

We performed a similar \vdep versus \Sigmol analysis for our suite of galaxy merger simulations. Since we are particularly interested in how the starburst activity influences GMC properties, we focus on the period right before and after the second passage where we can see the largest contrast in SFR. In Fig. \ref{fig:Sigmol_vdep_merger1} we show some example snapshots of \vdep versus \Sigmol for different merger stages during the second passage, along with \Sigmol and \alphavir maps at each snapshot. Note that the datacube is centered on the primary galaxy G3. At the time of first snapshot (2.54 Gyr), right before the start of the second perigalactic passage, the simulated mergers still have \Sigmol and \vdep that are similar the isolated galaxies. Then the molecular gas quickly transitions to a more turbulent state with much higher \vdep after the second passage along with a dramatic increase in global SFR, as shown in the snapshot for 2.66 Gyr (middle panel of Fig. \ref{fig:Sigmol_vdep_merger1}). The merger at this time still shows two separate nuclei in the zeroth moment map; this is similar to our observed mergers, the Antennae and NGC 3256. At this time, the \vdep versus \Sigmol contours for the simulated merger lie above the trend seen for the PHANGS galaxies, similar to NGC 3256, but in contrast to the Antennae, which still lies along the trend of the PHANGS galaxies. The larger deviation above the PHANGS trend implies higher \alphavir. We note that different \alphaco choices will affect the position of the contours. If we choose the ULIRG \alphaco instead of the Milky Way value, the Antennae would have \alphavir similar to that of NGC 3256 and our G2\&G3 merger. The uncertainty in the correct \alphaco value to use makes it difficult to interpret the data for the Antennae in this context.

The bottom panel of Fig. \ref{fig:Sigmol_vdep_merger1} shows the snapshot at 2.87 Gyr, which marks the post-merger stage after the final coalescence of two nuclei (defined here as the time at which the two central supermassive black holes are at a distance of 500 pc for the last time). This is the time when both \Sigmol and \vdep reach their highest values. We can see that most of the molecular gas is concentrated in the central 1 kpc region, with \Sigmol reaching 1000 \coldenunit. \vdep reaches 200 \velu, which is even higher than the \vdep observed in NGC 3256, which is in an earlier merging stage when the two nuclei have not yet coalesced. 

To better quantify the variation of \Sigmol and \vdep during the second passage, we plot the 16th, 50th and 84th percentile of the mass-weighted values for all pixels of each snapshot during the second passage in Fig. \ref{fig:Sigmol_vdep_deviation}. We also normalize both the median \Sigmol and \vdep to the median values of the isolated G3 galaxy at 0.625 Gyr (Fig. \ref{fig:Sigmol_vdep}) to show how the merging event affects the GMC properties during the second passage. Both \Sigmol and \vdep increase significantly during the merger, with a maximum increase of a factor of 10. The increase in \vdep and \Sigmol is roughly of the same order. Eq. \ref{eq: alphavir} shows that a constant \alphavir requires $\sigma_v^2 \propto \Sigma_{\mathrm{mol}}$. These results imply that our simulated merger will have higher \alphavir compared to PHANGS galaxies.

\subsection{The virial parameters of GMCs}
\label{sec:alphavir_measure}
\begin{figure*}[htb!]
	\gridline{
		\fig{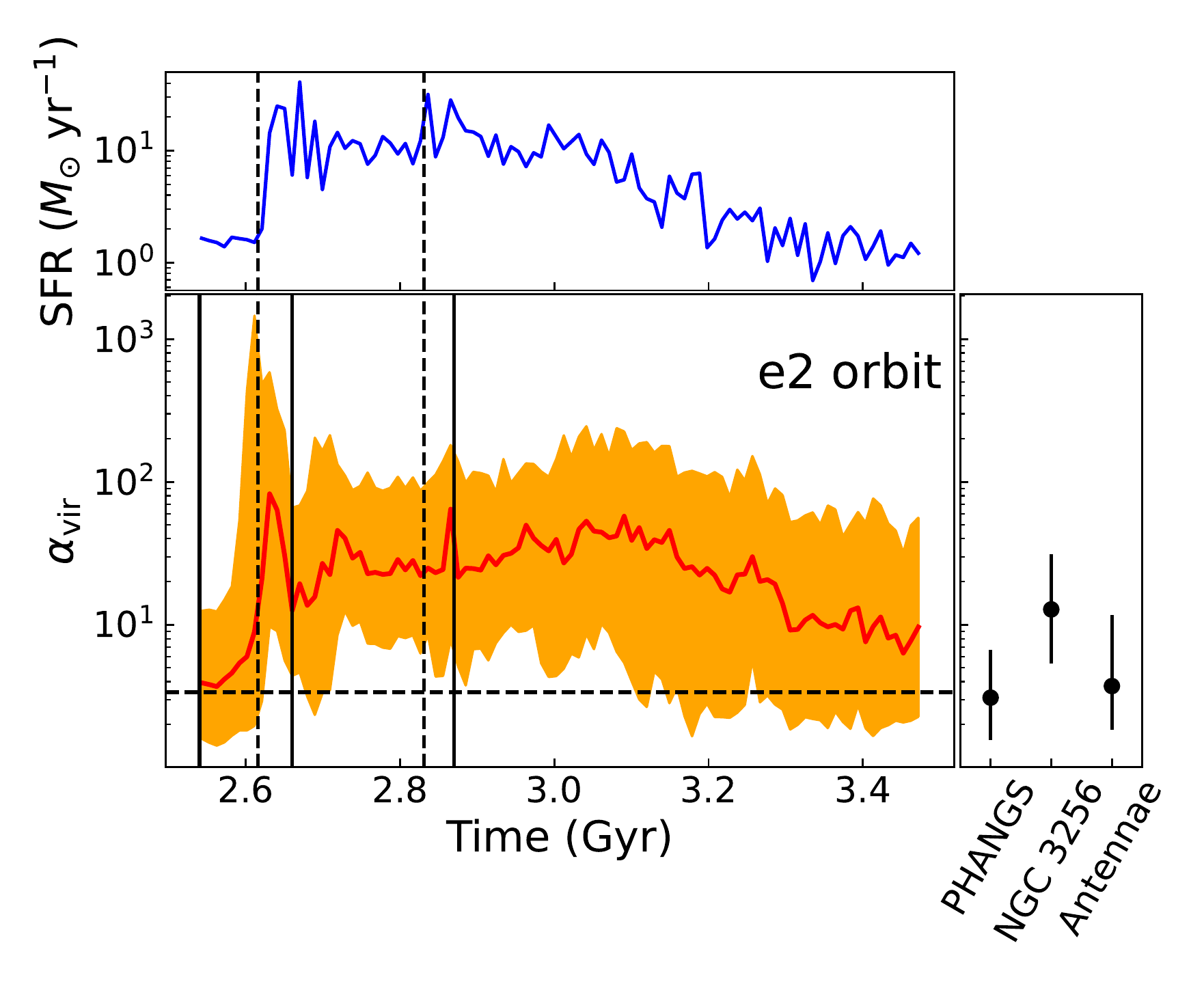}{0.5\textwidth}{}
		\fig{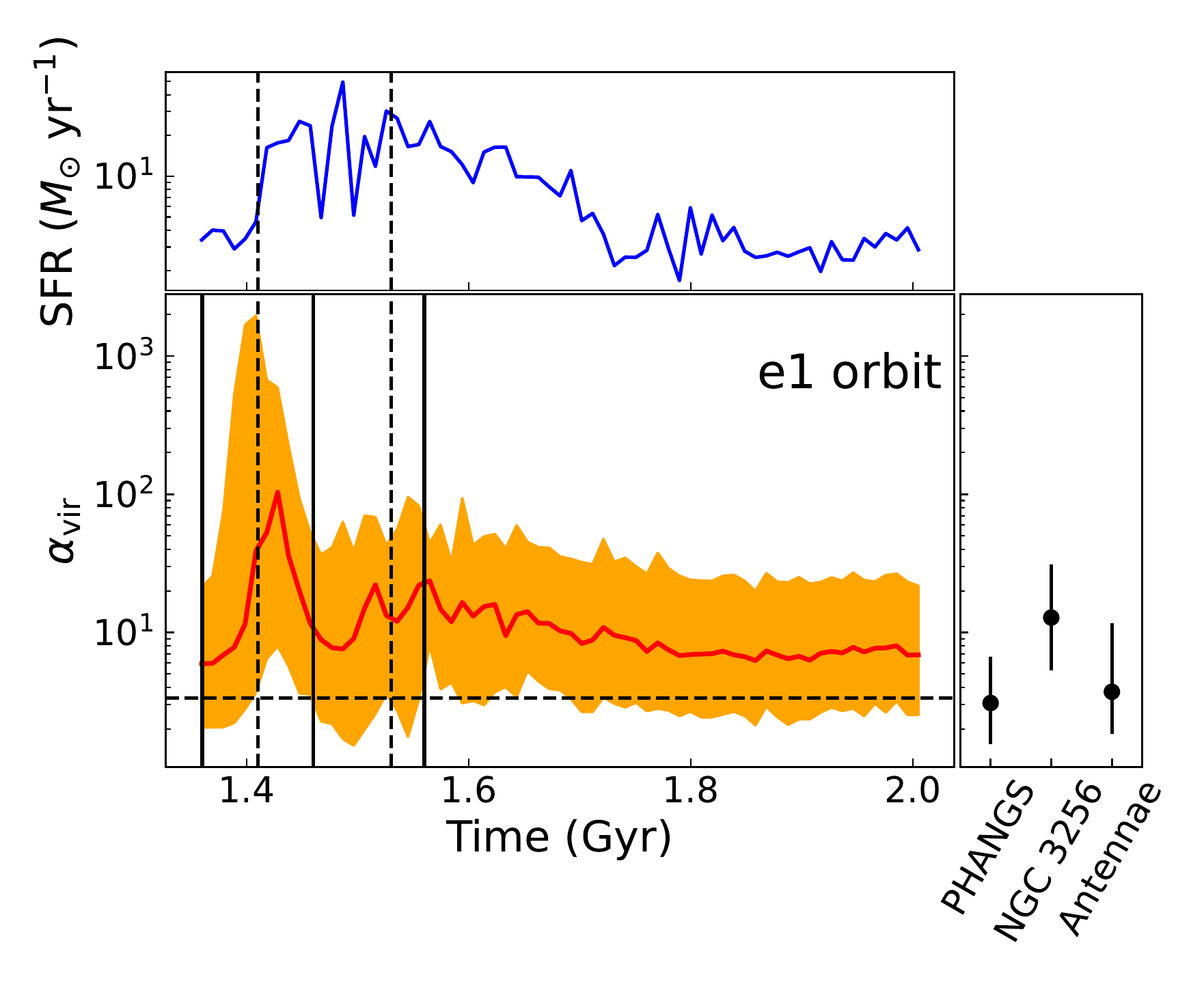}{0.5\textwidth}{}
	}
	\vspace{-2\baselineskip}
	\caption{\alphavir versus time for the G2\&G3 mergers in (left) the e2 orbit and (right) the e1 orbit viewed from `v0' angle during the final coalescence. (\textit{Left}) The red line is the mass-weighted median for \alphavir from the simulation. The orange shaded region includes data within the 16th and 84th quantile of \alphavir values. The dashed lines correspond to the start of the second passage and the final coalescence of the two nuclei. The three solid lines correspond to the merger times shown in Fig. \ref{fig:Sigmol_vdep_merger1}. The horizontal dashed line indicates the median \alphavir for the isolated G3 galaxy at 0.625 Gyr (Fig . \ref{fig:Sigmol_vdep}) as a baseline for comparison.  
	The upper panel shows SFR versus time for the second coalescence and the right panel shows the 16th, 50th and 84th quantile of \alphavir for PHANGS, NGC 3256 and the Antennae from the observations. In calculating \alphavir, we use the U/LIRG \alphaco for NGC 3256 and the Milky Way value for PHANGS and the Antennae. (\textit{Right}) Same plot for G2\&G3 merger in the `e1' orbit during the final coalescence. The 3 solid lines correspond to 3 snapshots in Fig. \ref{fig:Sigmol_vdep_e1}. 
	The `e1' orbit has a smaller impact parameter than the `e2' orbit. We can see the global \alphavir increases dramatically right after the second passage as SFR rises. The peak SFR also roughly corresponds with the peak \alphavir, which suggests the high \alphavir might be caused by the feedback from the starburst. }
	\label{fig:alphavir}
\end{figure*}  

During the second passage, we see that the \vdep vs \Sigmol distribution for our simulated merger lies above the trend observed for the PHANGS galaxies. A higher \vdep for a given \Sigmol  means the GMCs in these mergers are more turbulent and less gravitationally bound than in normal spiral galaxies. 

We adopt the same approach as in observations to calculate \alphavir for pixel-based GMC pixels using Eq. \ref{eq: alphavir}. Since the simulation data do not have a telescope ``beam" and each pixel in this analysis is treated as an independent GMC, we set $R$ to be half the size of each pixel (50 pc). With constant $R$, \alphavir depends only on \vdep and \Sigmol. Higher \vdep at a similar \Sigmol thus implies that \alphavir values for GMCs in simulated mergers are higher than the values for PHANGS or simulated isolated galaxies. Higher values for \alphavir are also found for NGC 3256 \citep{brunetti_highly_2020, brunetti_extreme_2022} and the Antennae \citep[][]{brunetti_cloud-scale_2022}. 

Fig. ~\ref{fig:alphavir} shows \alphavir as a function of time during the period near the second pericentric passage for the merger simulations with ``e2" and ``e1" orbits and viewed from ``v0" angle. \alphavir stays low before the second passage and suddenly rises after the passage along with a sudden increase in SFR. The peak of median \alphavir can reach $\sim$100. After the second passage, \alphavir gradually dies down as the SFR also decreases. During the entire merging process, we generally see a good correspondence between the SFR and \alphavir peaks, which suggests that the \alphavir value is either regulated by feedback from star formation or that both SFR and \alphavir increase together as a result of the merger.

\alphavir for our fiducial `e2' orbit is generally higher than that of the `e1' orbit and stays at higher values for a significantly longer time. The `e2' orbit has a higher impact parameter than the `e1' orbit (Section \ref{subsec:merger_suite}). Therefore, we would expect more gravitational potential energy transferred to the kinetic energy of individual GMCs, potentially making these GMCs less gravitationally bound. The \alphavir values for the `e1' orbit are more similar to the \alphavir of NGC 3256 and the Antennae and the `e1' orbit is more similar to the orbit of the Antennae. We note that both the Antennae and NGC 3256 are at the very start of their second passages \citep{privon_dynamical_2013,renaud_diversity_2019}. At this stage, there are significant variations in \alphavir, which makes it difficult to pick the exact snapshot that matches the observation. If we use the U/LIRG \alphaco instead of the Milky Way value, \alphavir for the Antennae would be similar to that of NGC 3256. We will discuss our \alphaco choices further in Section \ref{subsec: obs_simul_comparison}.

\subsection{Molecular Gas in the central 1 kpc region}
\label{subsec:gas_concentration}

\begin{figure}[htb!]
    \gridline{
        \fig{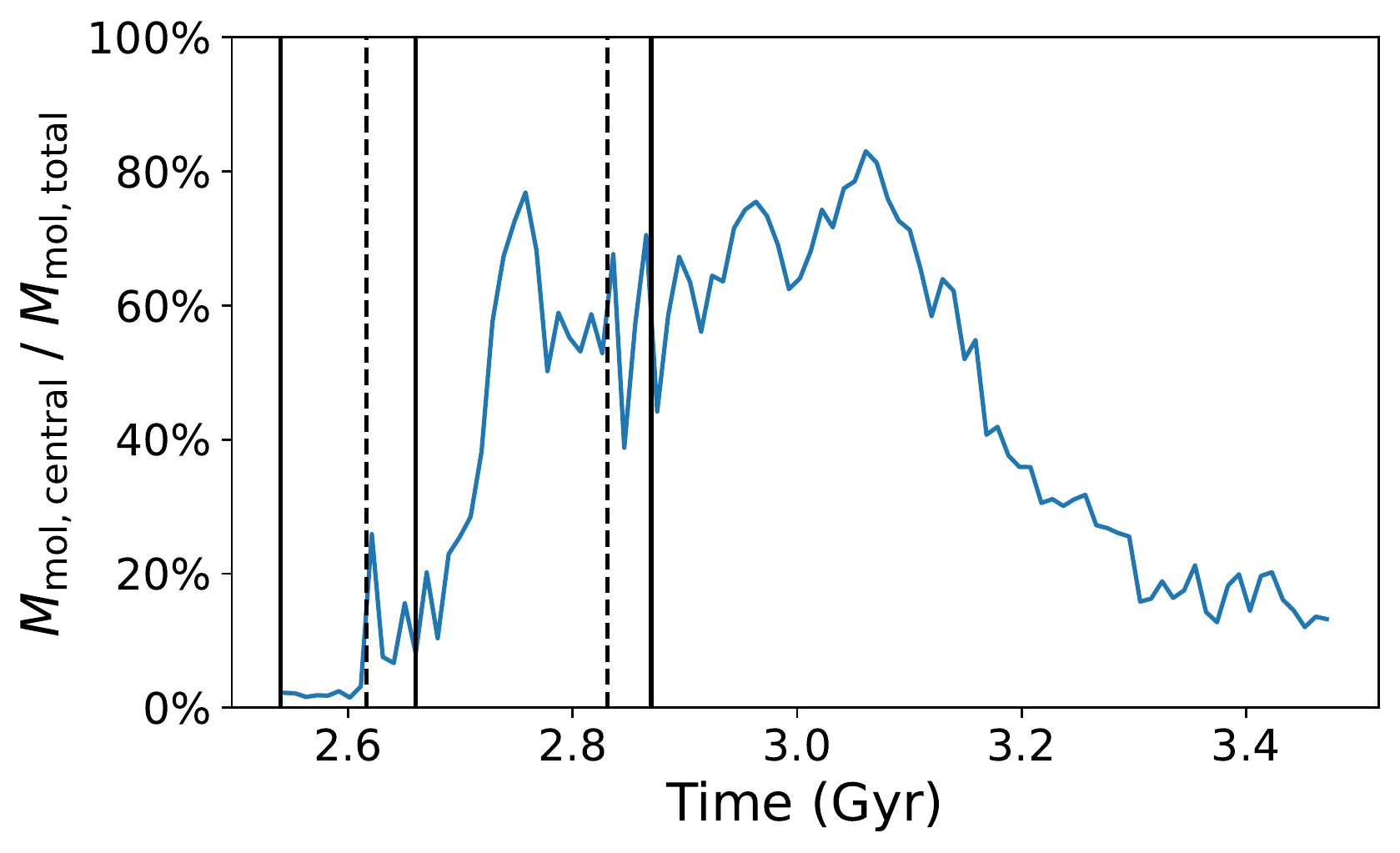}{0.45\textwidth}{}
        }
    \vspace{-0.4 in}
    \caption{The ratio between molecular gas mass within the central 1 kpc radius circle of the G3 galaxy and total molecular gas inside our FOV of 25 kpc. During the second coalescence between 2.7 Gyr and 3.2 Gyr, more than 50\% of molecular gas is concentrated within the central 1 kpc region, which indicates the \Sigmol increase we see in the simulated merger during the second passage is probably due to this gas concentration. }
    \label{fig:central_concentration}
\end{figure}

From the moment 0 maps in Fig. \ref{fig:Sigmol_vdep_merger1}, we can see that most molecular gas is concentrated in the center during the post-merger phase after 2.83 Gyr. This is consistent with the traditional scenario that the central starburst activity is caused by the inflow of molecular gas due to the loss of angular momentum \citep{hernquist_tidal_1989, barnes_fueling_1991, mihos_ultraluminous_1994, mihos_gasdynamics_1996, barnes1996, moreno2015}. To quantify how much of the molecular gas is concentrated in the center, Fig. \ref{fig:central_concentration} shows the molecular gas mass within the central 1 kpc, and the ratio between this value and total molecular gas mass. The fraction of molecular gas concentrated in the center reaches as high as 80\% for a significant period of time ($\sim 500$ Myr) around the final coalescence. On the other hand, \citet{moreno_interacting_2019} shows that the total molecular gas mass decreases during the second passage. Therefore, the overall high \Sigmol values of GMCs across our simulated merger compared to isolated galaxies are mostly due to the central gas concentration. 

\begin{figure*}[htb!]
    \centering
    \gridline{
        \fig{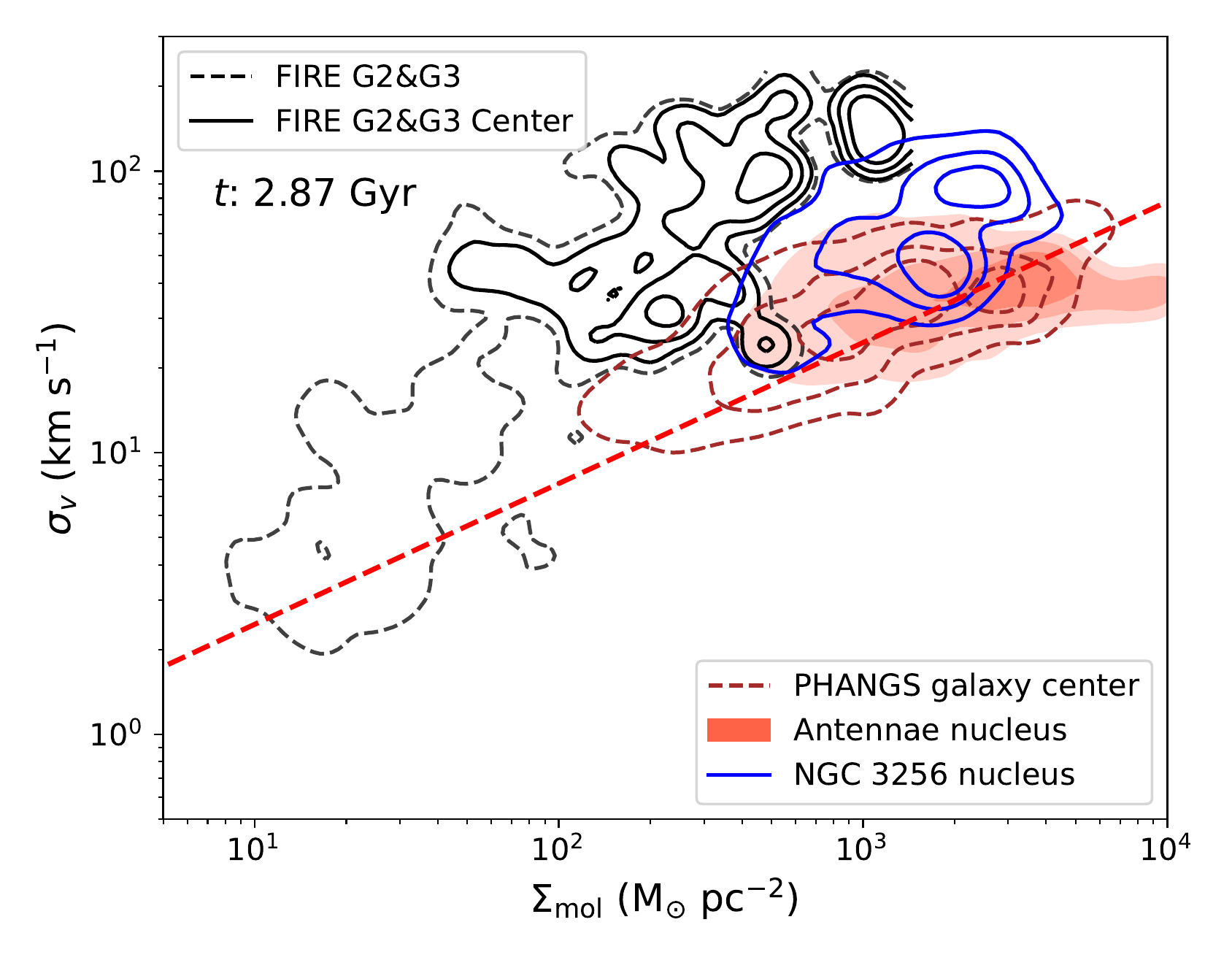}{0.5\textwidth}{}
        \fig{Figures/G2G3_alphavir_center.pdf}{0.5\textwidth}{}
        }
    \vspace{-0.4 in}
    \caption{ (\textit{Left}) The \vdep versus \Sigmol contour for the entire (dashed contour) and central 1 kpc region (solid contour) of the G2\&G3 merger at 2.87 Gyr viewed from the `v0' angle. We also show contours for the centers of PHANGS galaxies (brown dashed contours), the Antennae (orange shaded contours) and NGC 3256 (blue contours). We can see the central region in our simulated merger generally has the highest \vdep and \alphavir. (\textit{Right}) The mass weighted median \alphavir for molecular gas in the entire (blue) and central (orange) region of G2\&G3 merger viewed from `v0' angle. We see that \alphavir for the entire disk gradually settles back to the original low value, while that for the central region keeps a high value until the end of the simulation.
    }
    \label{fig:vdep_central}
\end{figure*}

Fig. \ref{fig:vdep_central} shows the \vdep versus \Sigmol distribution for pixels in the central kpc region of the G2\&G3 merger at 2.87 Gyr (red aperture in Fig. \ref{fig:Sigmol_vdep_merger1}), along with pixels in the center of PHANGS galaxies, the Antennae and NGC 3256. We can see the pixels in the center of the G2\&G3 merger have a larger deviation from the PHANGS trend than NGC 3256, which indicates that the G2\&G3 merger has GMCs with larger \alphavir in the center. We also show the mass weighted median \alphavir for the entire and central region of G2\&G3 merger as a function of time (Fig. \ref{fig:vdep_central} right). \alphavir in the center is generally higher than for the entire region, which indicates that GMCs in the center are more perturbed and less gravitationally bound. At the time right after the second passage, we see dramatic fluctuations of \alphavir for both the center and the entire galaxy, which is probably due to the complex and constantly varying gas morphology during this period. Moreover, we might see two GMCs that are far apart in 3D space but lie along the same line of sight, which cause large measured \alphavir value, but in a short time they no longer lie along the same line of sight, which causes a sudden drop of \alphavir. At the post-merger phase, \alphavir values are more stable.  
However, we see that \alphavir of the disk region gradually settles down while the central \alphavir keeps increasing. This might indicate that GMCs in the central region take more time to settle down to their normal states, which may be due to the starburst activity in the center. We also see high \alphavir for the center at the very start (2.54 Gyr), which probably means GMCs in the center at this time have not recovered from the starburst event that occured during the first peri-galactic passage. 



\subsection{Correlation between the central SFR and GMC Properties}
\label{subsec:SFR_GMC_correlation}

\begin{figure*}[htb!]
	\gridline{
		\fig{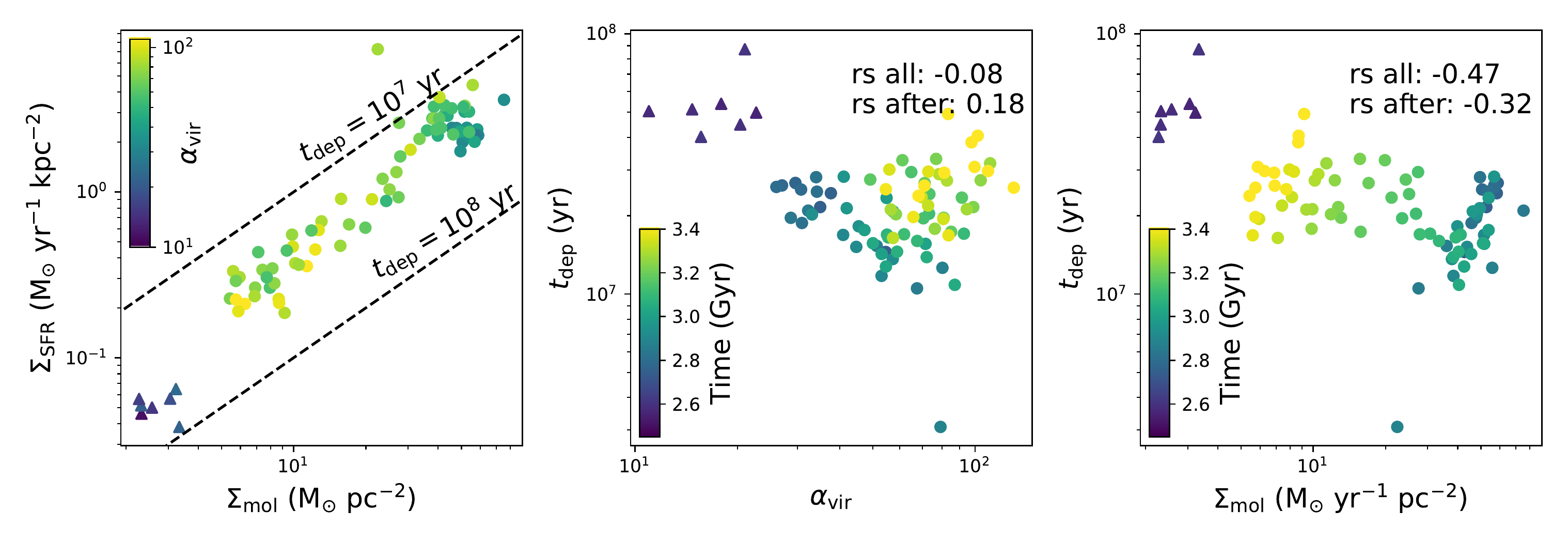}{\textwidth}{}
	}
	\vspace{-2.5\baselineskip}
	\caption{(\textit{Left}) SFR surface density \SigSFR versus \Sigmol color coded by the mass-weighted median \alphavir for the central 1 kpc region of the the G2\&G3 merger during the second passage of the `e2' orbit viewed from `v0' orientation. We include simulated data points within 2.54 -- 2.61 Gyr (before the second passage; trangle) and 2.73 -- 3.47 Gyr (after the second passage; circle). Both \SigSFR and \Sigmol are calculated as total central SFR or \Mmol within 1 kpc radius divided by the aperture size, while \alphavir is the mass weighted median of pixels inside the aperture. The two dashed line indicate constant depletion times ($t_{\mathrm{dep}}=\Sigma_{\mathrm{mol}}/\Sigma_{\mathrm{SFR}}$) of 10$^7$ and 10$^8$ years. (\textit{Middle}) \tdep versus \alphavir for the central 1 kpc. (\textit{Right}) \tdep versus \Sigmol for the central region. The label ``rs all" shows the Spearman coefficient between \tdep and \alphavir and \Sigmol for all data points while ``rs after" shows the Spearman coefficient only for data after the second passage. We can see there is no significant correlation between \alphavir and \tdep, which is against our expectation that low \alphavir clouds will consume molecular gas at faster rate. }
	\label{fig:tdep_SFR_alphavir}
\end{figure*}

The driving mechanism behind the SFR enhancement in mergers is of great interest to the study of star formation and galaxy evolution. One approach to tackle this problem is to decompose the SFR into the following 2 terms
\begin{equation}
    \mathrm{SFR} = \frac{M_{\mathrm{mol}}}{t_{\mathrm{dep}}}, 
\end{equation}
where $t_{\mathrm{dep}}$ is the depletion time, defined as the time for star formation to consume the available molecular gas. This approach makes it clearer that the rise in SFR could be either due to a larger amount of molecular gas ``fuel driven") or shorter depletion time (``efficiency driven"). The simulations \citep[e.g.][]{moreno2021} and observations \citep[e.g.][]{thorp_almaquest_2022} indicate that both terms contribute to the SFR enhancement at kpc scales. Moreover, many studies of the Kennicutt-Schmidt relation in U/LIRGs at kpc scales show that these starburst mergers have relatively short \tdep of $\sim 10^8$ yr compared to normal spiral galaxies of $\sim 10^9$ yr \citep[e.g.][]{daddi_different_2010}, which confirms the role of efficiency driving in mergers. With our simulations being able to probe the molecular gas at GMC scales, we can explore how \tdep is correlated with GMC populations in different regions. 

For this analysis, we focus on the molecular gas and star formation in the central 1 kpc region since most gas is concentrated here during the second passage (see Section \ref{subsec:gas_concentration}). We measure the mass-weighted median \alphavir in this central region as a metric for GMC dynamical state in the center. Fig. \ref{fig:tdep_SFR_alphavir} shows \Sigmol and \SigSFR color-coded by \alphavir for the central region as a function of time. We calculate the average \Sigmol and \SigSFR by dividing the total \Mmol or SFR in the central region by the aperture size. We show the data points within the period of 2.54 -- 2.61 Gyr (before the second passage) and 2.73 -- 3.47 Gyr (after the second passage) for comparison. We exclude the data points between the start of the second passage (2.62 Gyr) and the time when the central/total gas fraction starts to reach 50\% (2.73 Gyr) because data points from this period show a large deviation from the major trend in \SigSFR vs \Sigmol diagram. The large deviation is probably because the limited amount of molecular gas is highly perturbed in the central region. Gas is either quickly consumed without being replenished in time, or just concentrated and has not formed stars yet, which causes the large scatter in the \SigSFR vs \Sigmol  relation. On the other hand, before and after this period, the central region is in a relatively stable state when the molecular gas is constantly replenished to fuel star formation activity. 

In the left panel of Fig. \ref{fig:tdep_SFR_alphavir}, we can see that \tdep becomes shorter as \Sigmol and \SigSFR increase. The points at the lower left end of the \Sigmol correspond to the times before the second passage, which also have relatively low \alphavir. In contrast, the \alphavir after the second passage is significantly higher. We also note that \tdep even before the second passage ($\sim 10^8$ yr) is quite shorter than that of normal spiral galaxies ($10^9$ yr). The difference could be due to different dynamical timescales of simulated and observed galaxies. At this time, we can see \alphavir for the central region is already $\sim 10$ which indicates the molecular gas in the central region has already been disturbed. 

There is no significant correlation between \tdep and \alphavir, with Spearman coefficient of -0.08 for all data points and of 0.18 for data after the second passage, which is against our expectation that low \alphavir gas form stars more quickly. We can also clearly see a distinction between \alphavir before and after the second passage. The \alphavir before the second passage is relatively small and corresponds to larger \tdep while the \alphavir after the second passage is significantly larger but corresponds to shorter \tdep. This again is inconsistent with our expectation that low \alphavir GMCs should form stars more easily. Other physical mechanisms rather than self-gravity of individual GMCs may be needed to help the molecular gas to collapse (see detailed discussion in Section \ref{subsec:alphaVir_SFR}). 

On the other hand, we see an anti-correlation between \tdep and \Sigmol. This relation is similar to the global Kennicutt-Schmidt relation where the gas rich U/LIRGs have the shorter \tdep \citep{daddi_different_2010}. One explanation for this trend is that the fraction of dense gas (traced by HCN) that are actually forming stars (i.e. traces the self-gravitating gas fraction) is increasing as \Sigmol increases \citep[e.g. ][]{gao_star_2004,2023arXiv230106478B}. If we assume \Sigmol is proportional to the mean volume density of molecular gas in the central region, we would expect larger fraction of molecular gas above the dense gas threshold ($n > 10^4$ cm$^{-3}$) in FIRE-2 simulation \citep[][]{hopkins_new_2015}, which leads to faster star formation and shorter \tdep. 


\section{Discussion}

\subsection{How can high \alphavir gas form stars in simulated mergers?}
\label{subsec:alphaVir_SFR}

As shown in Section \ref{sec:alphavir_measure}, \alphavir generally stays above 10 during the second passage for the G2\&G3 merger. If we assume star formation occurs in individual GMCs and is driven by the collapse of the clouds due to self-gravity , we would expect star forming GMCs to have \alphavir below 1. The combination of high \alphavir values and starburst activity is inconsistent with this expectation, unless the velocity dispersion is being driven to higher values by infall motion. Furthermore, we find no correlation between \alphavir and \tdep (Section \ref{subsec:SFR_GMC_correlation}), which suggests low \alphavir values do not strongly affect the depletion time in our simulations. However, we need to note that our measurement of \alphavir from pixel-based method might not reflect the real \alphavir of individual GMC components, especially for the post-second-passage phase when molecular gas is concentrated in the center. Although observations \citep{brunetti_extreme_2022, sun_molecular_2020} show that cloud properties extracted from a pixel-based approach is generally consistent with the traditional cloud-based approach, they also show the pixel-based approach gives higher \vdep and \alphavir for molecular gas in galaxy centers. This is likely due to the superimposition of different GMC components along the same line of sight in gas-concentrated galaxy centers. \citet{sun_molecular_2022} find that \alphavir from pixel-based approach is $\sim$ 3 times higher than the cloud-based approach for galaxy centers. If we assume the same degree of overestimate in our simulation data for the merger center, we would expect the real \alphavir to be $\sim$ 10 during the second passage, still significantly higher than the critical value of 1 when clouds reach the self-collapsing criterion. We also note that even the observational cloud-based approach by extracting different GMC components from p-p-v data cube might still suffer from the projection effects. \citet{beaumont2013} find that \alphavir from p-p-p and p-p-v cubes have a factor of 2 difference for substructures in their cloud simulation due to a mismatch of substructures from these two data cubes. Therefore, one of our next steps is to perform cloud-finding algorithm \citep{2013ApJ...770..141B} on both p-p-p and p-p-v simulation data cubes to fully understand how GMCs evolve during the merging events. 


A possible explanation for large \alphavir is that GMCs that satisfy the self-collapsing criterion have already formed stars and become unbound or destroyed due to the stellar feedback. However, if this is the case, we would expect \alphavir to fluctuate around the critical value of 1. Furthermore, according to \citet{benincasa_live_2020}, GMCs with high \alphavir ($>$10) have significantly shorter lifetimes ($\sim$2 Myr) than GMCs with low \alphavir ($\sim$1; lifetime of $\sim$10 Myr). If we assume all GMCs are of the same population but at different evolutionary stages, we would expect GMCs to stay at low \alphavir state for a longer time and hence we should be more likely to catch these low \alphavir GMCs in our simulation snapshots. Instead, we see \alphavir constantly higher than 10 during the starburst activity (Fig. \ref{fig:alphavir}), which is inconsistent with this scenario. 

It is perhaps likely that the explanation is that these GMCs are experiencing compression from the large-scale gravitational potential. This compression could add additional potential energy to balance the kinetic energy. Furthermore, they can trigger inflow of gas into GMCs and bring radial velocity ($V_r$) component into our \vdep measurement. \citet{ganguly2022} find in their simulation that $v_r$ could be an important factor to produce high measured \alphavir clouds. For GMCs in normal spiral galaxies and galaxy pairs (e.g., M 51), \citet{meidt_model_2018} show that the large-scale stellar potential could be responsible for holding individual GMCs in energy equipartition state. Compared to galaxies in their study, the starburst mergers in our study undergo more dramatic morphological changes, which could generate complicated gravitational tidal fields. \citet{renaud_fully_2009} show in their simulation that major mergers can produce fully compressive tidal fields that concentrate molecular gas and trigger starburst activities. These compressive tidal fields are believed to be responsible for creating the off-nuclei gas concentration region in the ULIRG, Arp 220 \citep{downes_rotating_1998}. In our next step to test this scenario, we will need to calculate tidal deformation timescale \citep[as in][]{ganguly2022} for each individual GMC and compare it with GMC free-fall and crossing timescales to see how important the external tidal field is compared to GMC self-gravity. 

Another possible explanation is that molecular gas is smoothly distributed rather than clumped into individual GMCs during the starburst activities. If this is the case, the star formation is regulated by the entire molecular disk rather than individual GMC components \citep{2018MNRAS.477.2716K}. \citet{wilson_kennicuttschmidt_2019} propose that the star formation in U/LIRGs is regulated by the hydrodynamic pressure of the molecular disk with a constant scale height. In observation, one way to test the smoothness of gas distribution is by comparing average gas surface density at different observing resolutions \citep{leroy2017}. \citet{brunetti_highly_2020} show that molecular gas in the LIRG, NGC 3256, is smoothly distributed based on this method. For our simulated merger, gas might be smoothly distributed during the second passage when most gas is concentrated in the center (e.g. at 2.87 Gyr, Fig. \ref{fig:Sigmol_vdep_merger1}. We could test this scenario by changing the pixel size in our p-p-v cubes and compare the average gas surface densities in the central region at different pixel resolutions. 



\subsection{Comparison with observations}
\label{subsec: obs_simul_comparison}

As shown in Section \ref{sec:merger}, our simulated merger generally has lower \Sigmol and higher \vdep and \alphavir compared to the two observed mergers, the Antennae and NGC 3256. We note that this simulation is not set to match the exact condition of the observed mergers, so some discrepancy between observations and simulations would be expected. From the observational side, the biggest uncertainty that comes into the measurement is the value of \alphaco. As mentioned in Section \ref{subsec:GMCs_mergers}, if we adopt the ULIRG \alphaco instead of the Milky Way value for the Antennae, we would find the Antennae to have similar \Sigmol and \alphavir as NGC 3256. In contrast, if we assume an even smaller \alphaco  for NGC 3256, that might bring the contours of the observations further away from the PHANGS trend and hence more similar to the simulation contours. However, various LVG modelings \citep{papadopoulos_molecular_2012, harrington_turbulent_2021} show that local U/LIRGs and high-z starburst galaxies generally have \alphaco above 0.8 \alphacou. In fact, a recent study by \citet{dunne_dust_2022} concludes that these starburst galaxies might actually have \alphaco equal to the Milky Way value by cross-correlating the CO luminosity with dust and CI luminosity. Therefore, a factor of 3 discrepancy in \alphavir between simulated mergers and NGC 3256 is probably real rather than due to measurement uncertainties. 

For the comparison between observations and simulations, we also note that the two observed mergers are both in an early stage after the second passage since we can still identify two separate nuclei. In this stage,  \alphavir is quite time-sensitive and it is difficult to match the exact same stage between the simulated and observed galaxies. Therefore, it is possible that both NGC 3256 and Antennae are caught at a specific merger stage with a lower \alphavir (although in the case of NGC 3256, still enhanced relative to PHANGS galaxies). In comparison, \alphavir in the simulations is relatively stable in the post-merger stage. This stability suggests that a comparison between simulations and observations of post-merger galaxies could be a useful next step. Moreover, post-mergers have a rather simple morphology, which simplifies the task of making quantitative comparisons.

It would also be interesting to compare the simulation results with starburst galaxies at high redshift. Recent works \citep[e.g.][]{dessauges-zavadsky2019, mestric_exploring_2022} show that we can probe GMC-scale star-forming clumps in gravitationally lensed objects at high redshift. These star-forming clumps generally show a similar \alphavir to GMCs of normal spiral galaxies in our local Universe despite using different \alphaco, and therefore lower than what we see in the simulations. However, these high-z targets likely live in a completely different environment than our idealized mergers. Specifically, high-z galaxies tend to have a much higher gas fraction, and thus can form self-gravitating clumps with low \alphavir more easily \citep{fensch_role_2021}. 


\subsection{Comparison with other simulations}

In this work, we use the non-cosmological simulations from \citet{moreno_interacting_2019} to compare  GMC properties in mergers and normal spiral galaxies. Two major advantages of this simulation suite are that it has a resolution of 1.1 pc (which is much smaller than typical GMC sizes) and it can model the ISM down to low temperatures ($\sim$ 10 K), both of which allow  us to match the molecular gas in simulations with CO observations. Various cosmological simulations show that mergers are responsible for enhancing gas fractions and triggering starburst activity \citep[e.g.,]{scudder_galaxy_2015, knapen_interacting_2015, patton_galaxy_2013, martin_limited_2017, rodriguez_montero_mergers_2019, patton_interacting_2020}. However, these simulations can only model gas with temperatures down to 10$^4$ K and hence are incapable of capturing the turbulent multi-phase structure of the ISM. An alternative approach is to compare observations with cosmological zoom-in simulations, which allows for higher resolution, more realistic feedback star formation thresholds, and more realistic modeling of the multi-phase ISM. Various authors have explored GMC properties, mostly in Milky-Way-like galaxies \citep[e.g.][]{guedes_forming_2011,ceverino_radiative_2014,sawala_local_2014,benincasa_live_2020,orr2021}, and they generally reproduce the GMC mass function in our Milky Way. However, only a handful of work \citep[e.g.][]{rey2022} has been done for GMCs in mergers. Also, the Milky Way is identified as a green-valley galaxy \citep{mutch_mid-life_2011} with lower SFR than typical spiral galaxies in the local universe. Therefore, due to the lack of zoom-in cosmological simulations on local mergers, we have adopted idealized simulations for this study. Furthermore, idealized simulations allow us to compare GMCs of control galaxies with those of mergers to directly study the impact of the merging event. 

Several idealized simulations have been performed to study molecular gas and GMC properties in mergers. \citet{karl2013} perform a merger simulation closely matched to the Antennae and find a great match on CO distributions between simulation and observations, which suggests insufficient stellar feedback efficiencies in the Antennae. \citet{li_formation_2022} perform a study of GMCs and young massive star clusters (YMCs) in Antennae-like mergers. They find that GMC mass functions for mergers have similar power-law slopes to normal spirals during the second coalescence but with much higher mass values. 
\citet{narayanan_co-h2conversion_2011} compare the \alphaco in mergers and normal spiral galaxies and find that the low \alphaco in mergers is mostly due to the high temperature and \alphavir of GMCs in the merger. They predict there is a transition stage with \alphaco between U/LIRG and Milky Way values and that \alphavir is tightly anti-correlated with \alphaco. In contrast, \citet{renaud_three_2019} show that \alphaco values drop quickly during each coalescence between two galaxies. We find similar behavior for \alphavir during the second coalescence, which might imply a similar drop in \alphaco \citep{narayanan_co-h2conversion_2011}. 


\section{Conclusions}

We summarize our main conclusions below:

\begin{itemize}
	\item Our pixel-by-pixel analysis shows that the FIRE-2 simulation by \citet{moreno_interacting_2019} successfully reproduces the \vdep vs \Sigmol relation for GMC-scale pixels measured for galaxies in the PHANGS survey. 
	
	\item The simulated mergers show a significant increase in both \Sigmol and \vdep for GMC-pixels by a factor of 5 -- 10 during the second passage when SFR peaks, which brings these pixels above PHANGS-trend in the \vdep vs \Sigmol diagram. This may indicate GMCs in these mergers are less gravitationally bound. We quantify this deviation by the virial parameter \alphavir and find that our simulated mergers have \alphavir of 10 $-$ 100, which is even higher than the observed \alphavir in NGC 3256. However, this discrepancy could be partly due to the high impact parameter in the initial set-up of the simulated mergers. Furthermore, we see a good correspondence between the increase in SFR and \alphavir, which suggest either the starburst feedback is responsible for dispersing the gas or the correlation is in response to gas compression. 
	
	\item Our simulated mergers show a clear gas concentration in the center during the second passage, with up to 80\% of molecular gas in the central 1 kpc region. Therefore, the GMC-pixels in the central region tend to have the highest \Sigmol. We also find these pixels tend to have the highest \vdep and \alphavir, which could be caused by the starburst feedback and the inflow of gas. 
	
	\item We explore if \alphavir at GMC scales is responsible for the varying depletion time (\tdep) in observed mergers. While we do not find a significant correlation between \tdep and \alphavir, we see a clear distinction before (small \alphavir, long \tdep) and after (large \alphavir, short \tdep) the second passage. This could be due to projection effects (multiple GMCs along the same line of sight) during the second passage when most of the molecular gas is concentrated in the central 1 kpc region. The next step is to run a cloud-identification algorithm on the data to disentangle this factor. We also suspect there might be some other mechanism, such as the stellar potential and inflow of gas, that helps the GMCs in starburst mergers to collapse and form stars. We also find that \tdep has a significant anti-correlation with \Sigmol for the central region. This may be due to higher \Sigmol leading to a higher fraction of dense gas, which shortens \tdep. 
\end{itemize}

In the future, we would like to expand our comparison to more observed and simulated mergers. From the observational side, we need larger samples of galaxy mergers spanning different evolutionary stages in order to understand how GMCs evolve throughout the merging. In addition, it is easier to compare the observations with simulations in the post-merger stage since the morphology is simpler and easier to quantify. The ALMA archive contains $\sim$ 40 U/LIRGs with GMC resolution CO 2-1 observations that can be used to build a more complete sample of GMCs in mergers at different stages. From the simulation side, it would be helpful to have simulations that better match the observed galaxies. The Antennae has been widely studied and matched by non-cosmological simulations \citep[e.g.][]{renaud_diversity_2019,li_formation_2022} but NGC 3256 is less well studied. Besides comparing with these non-cosmological simulations, we could also compare observation with cosmological simulations, such as FIREBox \citep{feldmann_firebox_2022}, that include local mergers. \\

\vspace{3mm}
We thank Dr. Jiayi Sun for his help to access to PHANGS data and insightful discussions about the comparison between simulation and observation. We thank Dr. Nathan Brunetti for access to \cotwo image and GMC catalogs of the observed mergers in his paper. 

This work was carried out as part of the FIRE collaboration. The research of C.D.W. is supported by grants from the Natural Sciences and Engineering Research Council (NSERC) of Canada and the Canada Research Chairs program. The research of H.H. is partially supported by the New Technologies for Canadian Observatories, an NSERC-CREATE training program. CB gratefully acknowledges support from NSERC as part of their post-doctoral fellowship program
[PDF-546234-2020]. The computations in this paper were run on the Odyssey cluster supported by the FAS Division of Science, Research Computing Group at Harvard University. Support for JM is provided by the Hirsch Foundation, by the NSF (AST Award Number 1516374), and by the Harvard Institute for Theory and Computation, through their Visiting Scholars Program. B.B. acknowledges support from NSF grant AST-2009679. B.B. is grateful for the generous support of the David and Lucile Packard Foundation and Alfred P. Sloan Foundation. The Flatiron Institute is supported by the Simons Foundation. 

This paper makes use of the following ALMA data: ADS/JAO.ALMA \#2015.1.00714.S and ADS/JAO.ALMA \#2018.1.00272.S. ALMA is a partnership of ESO (representing its member states), NSF (USA), and NINS (Japan), together with NRC (Canada), MOST and ASIAA (Taiwan), and KASI (Republic of Korea), in cooperation with the Republic of Chile. The Joint ALMA Observatory is operated by ESO, AUI/
NRAO, and NAOJ. The National Radio Astronomy Observatory is a facility of the National Science Foundation operated under cooperative agreement by Associated Universities, Inc.
\vspace{3 mm}
\facilities{ALMA}
\software{astropy \citep{the_astropy_collaboration_astropy_2013}, Spectral-Cube \citep{ginsburg_radio-astro-toolsspectral-cube_2019}}

\section*{Data Availability}

The datasets generated during and/or analysed during the current study are available from the corresponding author on reasonable request, contingent on approval by the FIRE Collaboration on a case-by-case basis.



\bibliographystyle{aasjournal}
\bibliography{references}{}


\restartappendixnumbering
\appendix

\section{Influence from different viewing angles}
\begin{figure}[htb!]
	\gridline{
		\fig{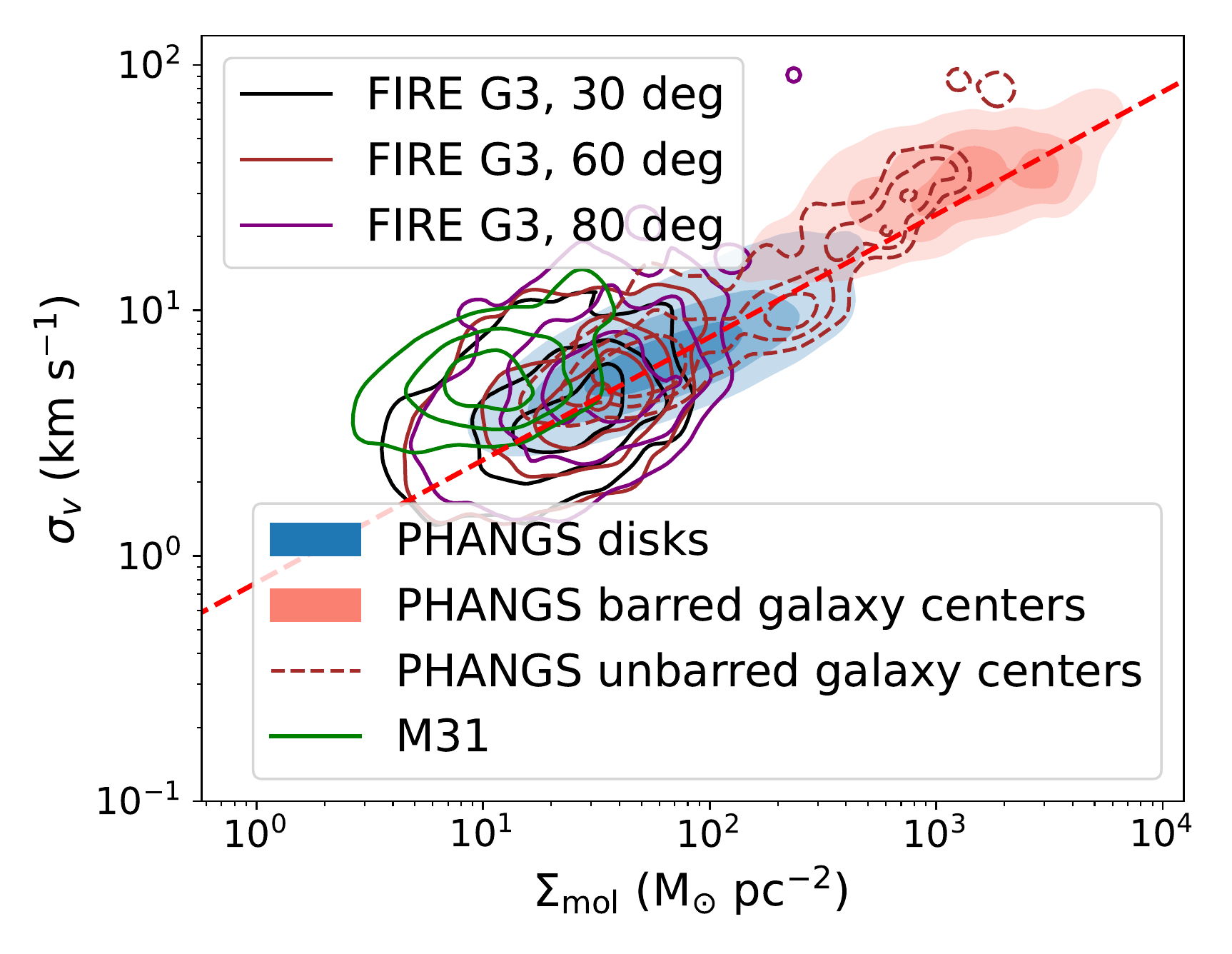}{0.45\textwidth}{}
	}
	\vspace{-0.4 in}
	\caption{Similar to Fig. \ref{fig:Sigmol_vdep} but with the simulated G3 galaxy viewed at different inclination angles (30, 60 and 80 degrees). }
	\label{fig:Sigmol_vdep_incls}
	\vspace{-1\baselineskip}
\end{figure}

One important factor that might influence \Sigmol measured from the simulations is the inclination angle at which the galaxy is viewed. For the simulated control galaxies, we pick the inclination angle of 30 degrees in Fig. \ref{fig:Sigmol_vdep}. By increasing inclination angle we might see significant increase in \Sigmol. Fig. ~\ref{fig:Sigmol_vdep_incls} shows the data for the G3 galaxy viewed with inclination angles of 30, 60 and 80 degrees. We see little increase in \Sigmol even for an inclination of 80 degrees. This is consistent with our expectation that individual clouds are resolved in the simulated data. For resolved spherical clouds, the observed surface density should always be the same despite different viewing angles.

\begin{figure}[th]
	\gridline{
		\fig{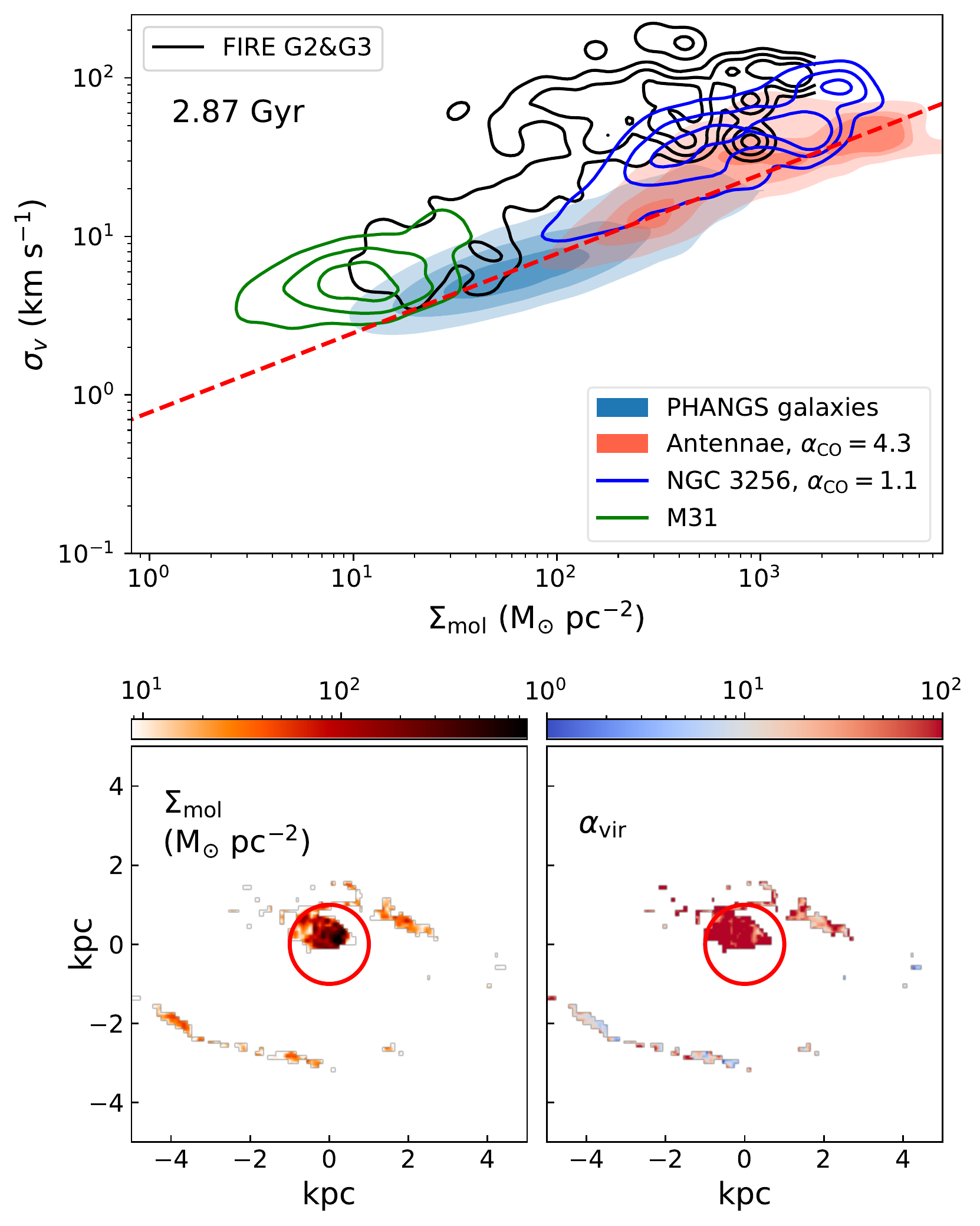}{0.45\textwidth}{}
	}
	\vspace{-0.4 in}
	\caption{The snapshot of FIRE-2 merger at a time of 2.87 Gyr with viewing angles of 'v1', which is roughly perpendicular (with an angle of 109 degree) to the 'v0' angle. (Upper) the \vdep vs \Sigmol distribution of GMCs. (Lower) The \Sigmol map of the snapshot. We can see that viewing from different angles still give us high \vdep measurement, which also suggests that \vdep we measure is not the velocity dispersion among GMCs along the line of sight. (lower) The \Sigmol and \vdep for this snapshot from 'v1' angle.  }
	\label{fig:Sigmol_vdep_merger2}
\end{figure}

For the simulated mergers, the molecular gas structure is more complicated than a single layer of gas disk. In this case, we might pick a specific angle where multiple clouds happen to lie along the same line of sight, which gives us large \vdep values. To test if this is the actual case, we examine a snapshot at 2.87 when we reach maximal \Sigmol and \vdep from a different angle ('v1'). As shown in Fig. \ref{fig:Sigmol_vdep_merger2}, we see similar gas distribution contour in \vdep vs \Sigmol contour. This along with the velocity spectrum in Fig. \ref{fig:vdep_central} suggests that the large \vdep and \alphavir we measured is intrinsic properties of individual GMCs. 

\section{Global gas fraction}
\label{sec:gas_fraction}

On global scales, we compare the molecular gas mass and total gas mass (including HI) of the FIRE-2 galaxies with observed values for normal spiral galaxies. We compare to both the  PHANGS galaxies \citep{leroy2021} as well as to the global gas properties from the xCOLDGASS survey, to confirm that the PHANGS galaxies are representative of star forming main sequence galaxies in our local universe. For xCOLDGASS, the molecular gas mass is extracted from \citet{saintonge_xcold_2017} and the total gas mass is from \citet{catinella_xgass_2018}.

\begin{figure*}[htb!]
	\gridline{
		\fig{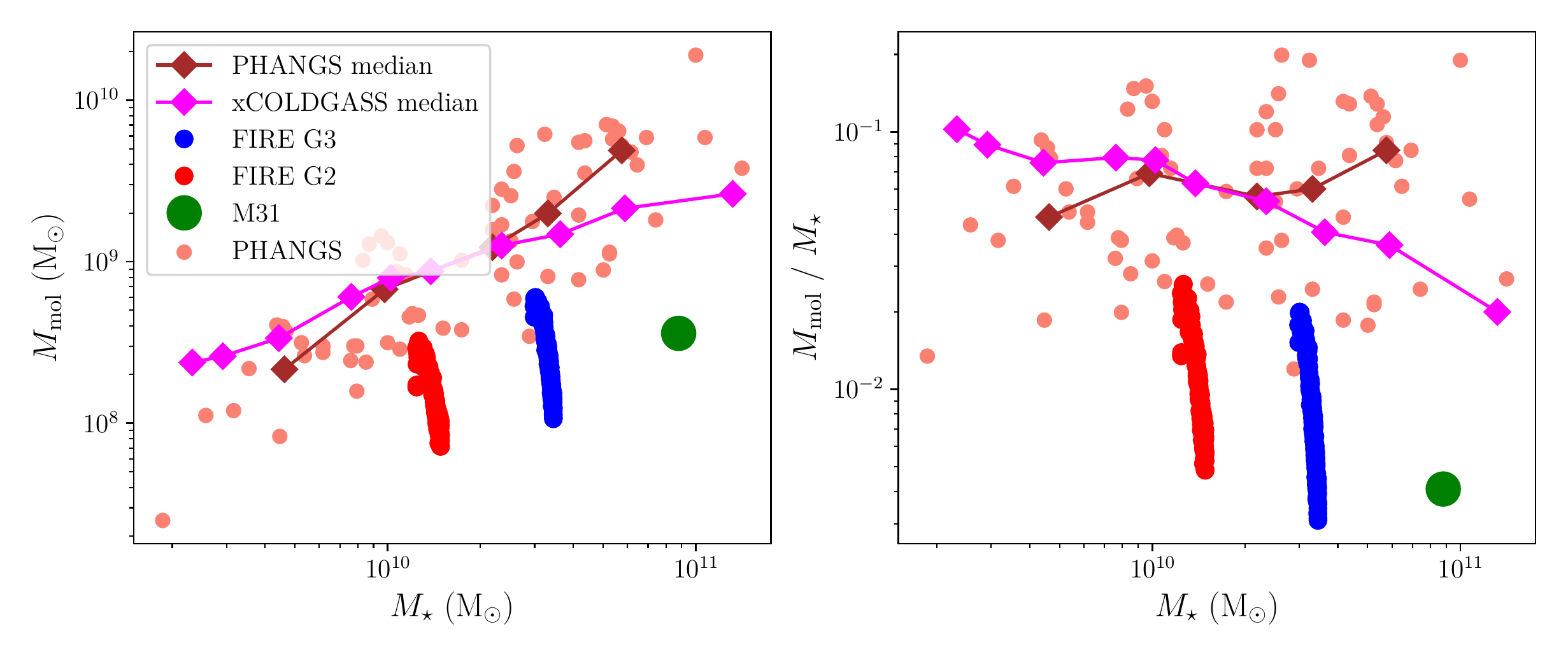}{0.8\textwidth}{}
	}
	\vspace{-0.5 in}
	\gridline{
		\fig{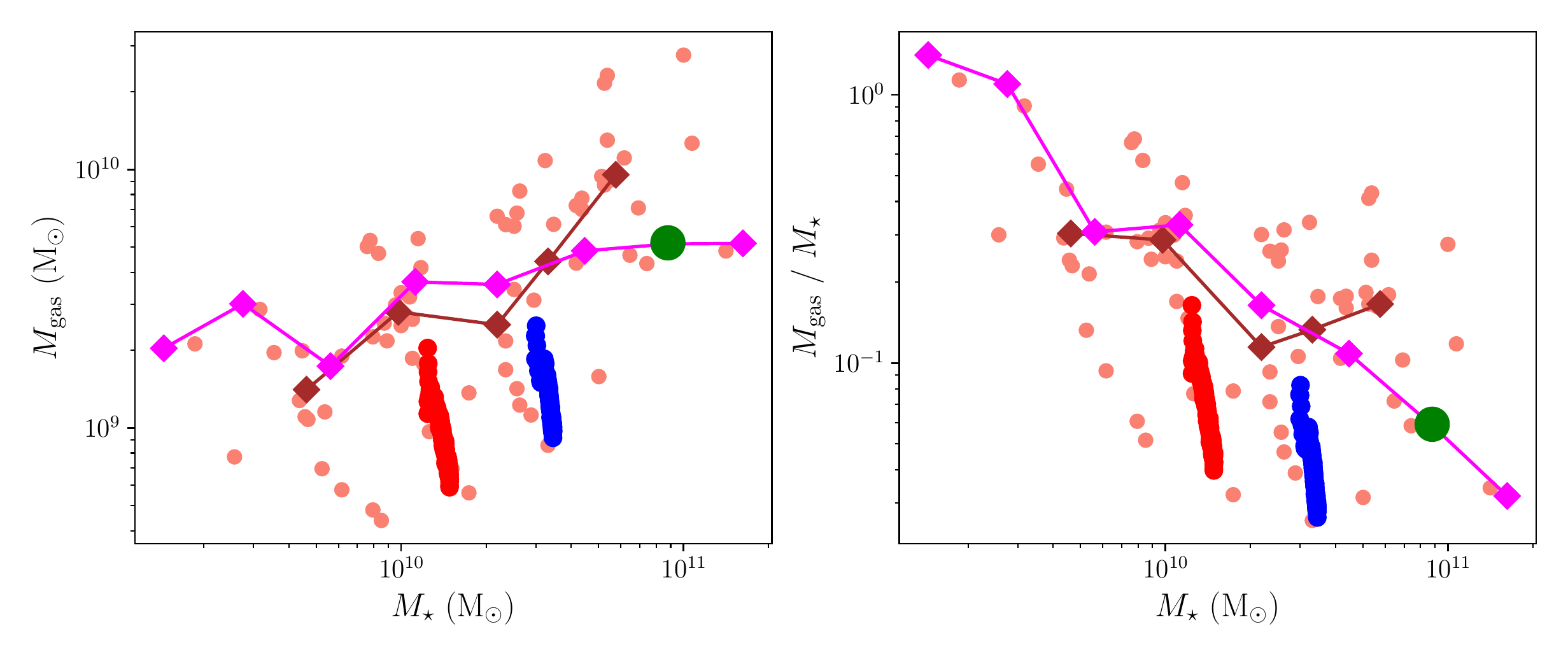}{0.8\textwidth}{}
	}
	\vspace{-0.4 in}
	\caption{(Upper Left) \Mmol versus \Mstar for PHANGS galaxies \citep[salmon dots;][]{leroy2021}, M 31 \citep[green filled circle; ][]{nieten_molecular_2006} and the G2 (red points) and G3 (blue points) simulated galaxies at different times in their evolution. Note that the G2 and G3 simulated galaxies lie significantly below  the star-forming main sequence defined by the xCOLDGASS sample. (Upper Right) \fmol versus \Mstar for the same galaxies. The molecular gas fractions of G2 and G3 are significantly lower than most of the PHANGS spiral galaxies. (Lower left) total gas versus \Mstar. (Lower right) total gas fraction versus \Mstar. These comparisons suggest that the low \Sigmol measured for the simulated galaxies might be due to the low total and molecular gas fraction in the initial set-up.  }
	\label{fig:Mol_fraction}
\end{figure*}

In interpreting the lower \Sigmol values seen in Fig. ~\ref{fig:Sigmol_vdep}, one possibility is that there may not be as much gas available to form high surface density clouds in the two simulated galaxies compared to the  PHANGS galaxies. 
Fig. \ref{fig:Mol_fraction} compares the global molecular gas masses, \Mmol, and molecular gas fractions, \fmol = \Mmol ~/ \Mstar, for the FIRE-2 mergers with those of the PHANGS galaxies from \citet{sun_molecular_2020}. We also show the median value of \Mmol and \fgas in each \Mstar bin for the PHANGS galaxies, as well as 
the weighted median of \Mstar and \fmol for galaxies in xCOLDGASS sample \citep{saintonge_xcold_2017}. The two median values are quite close to each other for galaxies with \Mstar of 10$^{9.5}$ -- 10$^{11}$ \solarmass, although the PHANGS galaxies seem to deviate somewhat from the xCOLDGASS sample in the highest and lowest mass bins. In contrast, the G2 and G3 galaxies both have \fmol $\sim$ 3 times lower than  typical PHANGS or xCOLDGASS galaxies of the same stellar mass.
Therefore, the small global \fmol may be responsible for producing the low \Sigmol values seen in the simulated galaxies. 

The low values of \fmol could be produced either by the initial set-up of the simulations or by physical mechanisms in the simulation that lead to inefficient conversion of gas into the cold phase. We can distinguish between these two options by calculating the total gas fraction \fgas including both HI and H$_2$. The lower panel of Fig. \ref{fig:Mol_fraction} shows the median of \Mgas and \fgas for the PHANGS galaxies and xGASS-CO samples \citep{catinella_xgass_2018} compared to the two simulated galaxies. The values of \fgas for both simulated galaxies are still $\sim$ 3 times lower than those of typical spiral galaxies with similar \Mstar. Therefore, it seems most likely that the low cold gas fraction, \fmol, is produced by a low total (cold+warm+hot) gas mass in the initial set-up of the simulations.

\section{Snapshots for `e1' orbit}
\label{sec:e1_orbit}

\begin{figure*}[h!]
    \gridline{
        \fig{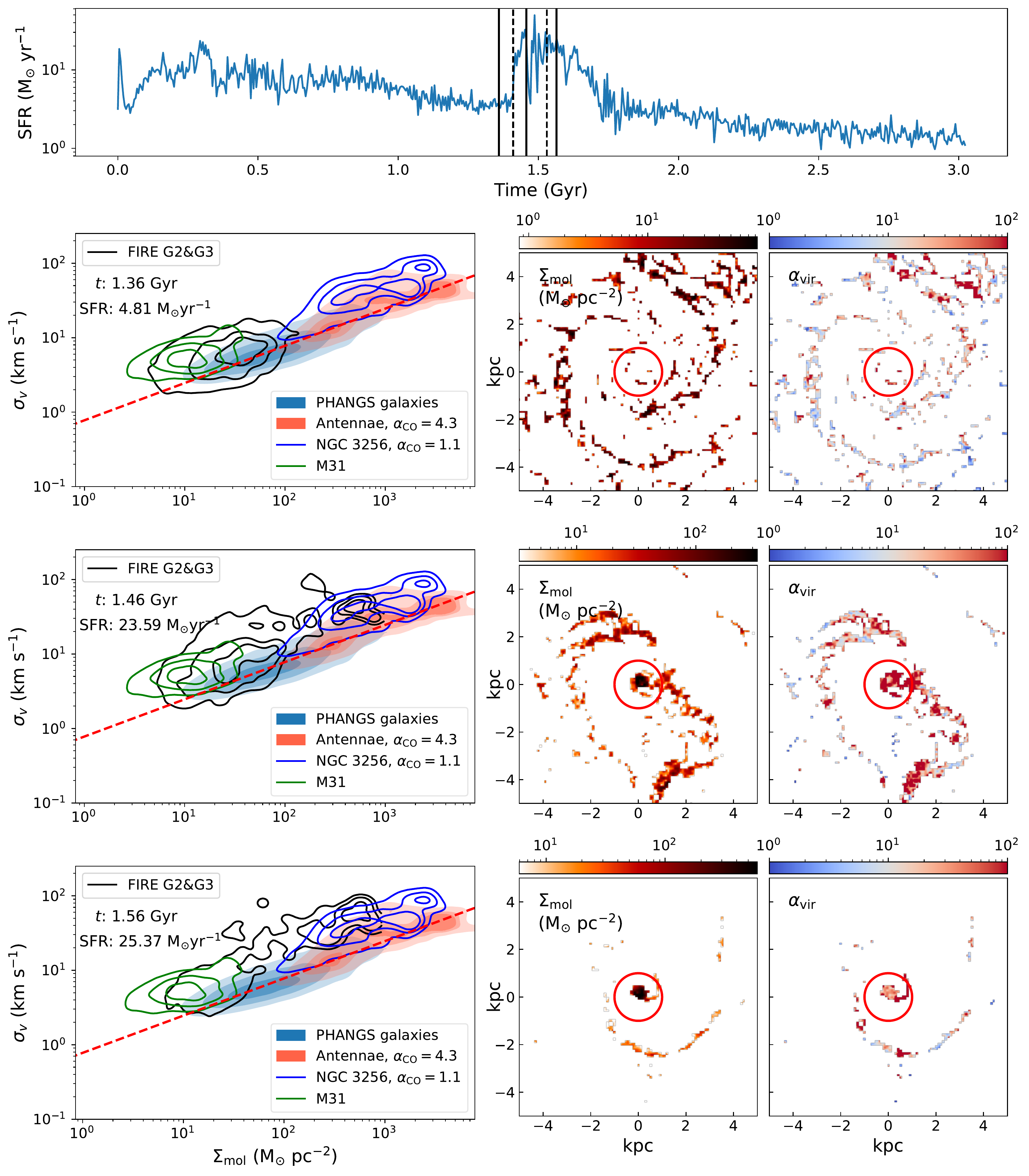}{0.95\textwidth}{}
        }
	\vspace{-0.2 in}
	\caption[C1]{Similar plot as Figure \ref{fig:Sigmol_vdep_merger1} but with SFR history and 3 snapshots from `e1' orbit. The interactive version of the animation is available at \url{https://heh15.github.io/merger_animation/G2G3_e1_v0_final}.}
	\label{fig:Sigmol_vdep_e1}
\end{figure*}

Here we show the SFR history and 3 example snapshots for G2\&G3 `e1' orbit in Fig. \ref{fig:Sigmol_vdep_e1}. 





\end{document}